%% file: main.tex
\documentclass[fleqn,usenatbib]{mnras}

\usepackage{newtxtext,newtxmath}

\usepackage[T1]{fontenc}
\usepackage{amsmath}
\usepackage{bm}
\usepackage{comment}
\usepackage{xcolor}

\newcommand{\p}{\partial}
\newcommand{\vel}{{\rm \bf V}}
\newcommand{\jj}{{\rm \bf J}}
\newcommand{\bb}{{\rm \bf B}}

\newcommand{\del}{\nabla}
\newcommand{\ptot}{\rm P^* }

\DeclareRobustCommand{\VAN}[3]{#2}
\let\VANthebibliography\thebibliography
\def\thebibliography{\DeclareRobustCommand{\VAN}[3]{##3}\VANthebibliography}

\usepackage{graphicx}	
\usepackage{amsmath}	

\title[RWI in 3D Non-Ideal MHD Disks]
{Rossby Wave Instability and Substructure Formation in 3D Non-Ideal MHD Wind-Launching Disks}

\author[C. -Y. Hsu et al.]{
Chun-Yen Hsu,$^{1}$\thanks{E-mail: kdj8qp@virginia.edu}
Zhi-Yun Li,$^{1}$
Yisheng Tu,$^{1}$
Xiao Hu,$^{1,2}$
Min-Kai Lin,$^{3,4}$
\\
$^{1}$Department of Astronomy, University of Virginia, Charlottesville, VA 22904, USA\\
$^{2}$Department of Astronomy, University of Florida, Gainesville, FL 32608, USA\\
$^{3}$Institute of Astronomy and Astrophysics, Academia Sinica, Taipei 10617, Taiwan\\
$^{4}$Physics Division, National Center for Theoretical Sciences, Taipei 10617, Taiwan
}

\date{Accepted XXX. Received YYY; in original form ZZZ}

\pubyear{2015}

\begin{document}
\label{firstpage}
\pagerange{\pageref{firstpage}--\pageref{lastpage}}
\maketitle

\begin{abstract}

Rings and gaps are routinely observed in the dust continuum emission of protoplanetary discs (PPDs). How they form and evolve remains debated. Previous studies have demonstrated the possibility of spontaneous gas rings and gaps formation in wind-launching disks. Here, we show that such gas substructures are unstable to the Rossby Wave Instability (RWI) through numerical simulations. Specifically, 
shorter wavelength azimuthal modes develop earlier, and longer wavelength ones dominate later, forming elongated (arc-like) anti-cyclonic vortices in the rings and (strongly magnetized) cyclonic vortices in the gaps that persist until the end of the simulation. Highly elongated vortices with aspect ratios of 10 or more are found to decay with time in our non-ideal MHD simulation, in contrast with the hydro case. This difference could be caused by magnetically induced motions, particularly strong meridional circulations with large values of the azimuthal component of the vorticity, which may be incompatible with the columnar structure preferred by vortices. The cyclonic and anti-cyclonic RWI vortices saturate at moderate levels, modifying but not destroying the rings and gaps in the radial gas distribution of the disk. In particular, they do not shut off the poloidal magnetic flux accumulation in low-density regions and the characteristic meridional flow patterns that are crucial to the ring and gap formation in wind-launching disks. Nevertheless, the RWI and their associated vortices open up the possibility of producing non-axisymmetric dust features observed in a small fraction of protoplanetary disks through non-ideal MHD, although detailed dust treatment is needed to explore this possibility. 

\end{abstract}




\section{Introduction}
\label{sec:introduction}
\input{Introduction.tex}


\section{Simulation setup} \label{sec:simultation}
\input{Simulation.tex}
\section{Non-ideal MHD Effects} 
\label{sec:non-ideal_coff}
\subsection{Chemical network for charge abundances}
\label{subsec:chemistry}
\input{Chemistry.tex}

\subsection{Non-ideal MHD Coefficients} \label{subsec:non-ideal}
\input{nonideal_MHD.tex}
\section{Formation of Gas Rings and Gaps in 2D and Their Stability in 3D} \label{sec:results}
\input{Result.tex}


\section{Conclusion} \label{sec:conclusion}

\input{Conclusion.tex}

\section*{Acknowledgements}
We thank the referee, Dr. Cui, Can, for a prompt and constructive report.
CYH acknowledges support from NASA SOFIA grant SOF-10\_0505 and NRAO ALMA Student Observing Support (SOS) and computing resources from UVA research computing (RIVANNA) and NASA High-Performance Computing. ZYL is supported in part by NASA 80NSSC20K0533 and NSF AST-2307199.
MKL is supported by the National Science and Technology Council (grants 111-2112- M-001-062-, 112-2112-M-001-064-, 111-2124-M-002-013- , 112-2124-M-002-003-) and an Academia Sinica Career Development Award (AS-CDA110-M06).
\section*{Data Availability}

The data underlying this article will be shared on reasonable request to the corresponding author.



\bibliographystyle{mnras}
\bibliography{ref}




\appendix

\section{Charge Abundances and Elsasser Numbers} \label{sec:appendix_a_chemical}
\input{appendix_Chemical_models}

\section{Rate Coefficients for Collisional Momentum Transfer} \label{sec:momentum_transfer_rate_coefficients}

Three charged species (ions, electrons, and grains) collide with the molecular hydrogen in our model. We take the collisional momentum transfer rate coefficients from \cite{Pinto08} and \cite{grassi19}. 
Note that the units of $R_{e, n}$,  $R_{i, n}$, and $R_{g, n}$ in this subsection are ${\rm cm^3s^{-1}}$, corresponding to the $R_{j, n} (T)$ in the Eq. (\ref{eq:beta_j_n}), so the Hall parameter is dimensionless.


For the collisional rate between electrons and molecular hydrogen, we use the rate
\begin{align}
    & R_{e, n} (T) = 10^{-9} \sqrt{T} \left[0.535 + 0.203 {\rm log_{10}(T)} - 0.163 [{\rm log_{10}(T)}]^2 \right]. \label{eq:rate_coef_e_n}
\end{align}
The collisional rate between ions and molecular hydrogen is given as
\begin{align}
    & R_{i, n} (T) = 2.21\pi  \sqrt{\frac{\alpha_{\rm pol} e^2}{m_{\rm red}}}. \label{eq:rate_coef_i_n}
\end{align}
where $\alpha_{\rm pol} = 8.06 \times 10^{-25} {\rm cm^3}$ is the polarizability of molecular hydrogen. 

The momentum transfer through collisions between charged grains and neutral hydrogen is 
\begin{align}
 & \langle R_{g, n} (T, Z) \rangle =2.21\pi  \sqrt{\frac{\alpha_{\rm pol} |Z| e^2}{m_{\rm red}}} \frac{a^{p+1}_{\rm c} - a^{p+1}_{\rm min}}{a^{p+1}_{\rm max} - a^{p+1}_{\rm min}} \notag \\
 & + \left( \frac{8 k_{\rm B} T}{\pi m_{\rm r}}\right)^{1/2} \frac{4 \pi \delta_{\rm L}}{3} \frac{a^{p+3}_{\rm max} - a^{p+3}_{\rm c}}{a^{p+1}_{\rm max} - a^{p+1}_{\rm min}} \frac{p+1}{p+3}, \label{eq:rate_coef_g_n} 
\end{align}
where 
\begin{align}
 & a_c(T,Z) = \frac{0.206}{\sqrt{\delta_{\rm L}}} \left( \frac{\alpha_{\rm pol}|Z|}{T}\right)^{1/4} \label{eq:critical_size} 
\end{align}
is a critical grain size to measure microscope grain-neutral reaction for the hard sphere and the Langevin rates \citep{Pinto08}. We set $\delta_{\rm L} = 1.3$ as recommended by \cite{Liu03}. 


\bsp	
\label{lastpage}
\end{document}

%% file: Introduction.tex
The detection of substructures is a watershed event in protoplanetary disk research \citep[PPDs, e.g.,][]{ALMA15, Andrews18}. While axisymmetric rings, gaps, and cavities are the most common features, non-axisymmetric structures such as spiral arms, lobes, and arcs are also observed. A particular type of non-axisymmetric feature is vortices, which are detected in, e.g., HD 142527 using CO emission lines \citep{Boehler21} and possibly through near-infrared scattered light \citep{Marr22}. Several mechanisms are proposed to produce the substructures, including planet-disc interactions \citep[e.g.,][]{Dong15,Bae17}, sintering of volatile ices outside snow lines \citep{Okuzumi16}, sharp change of disk properties at the dead zone boundary \citep[e.g.,][]{Flock15, Ruge16},
secular gravitational instability \citep[e.g.,][]{Takahashi16, Tominaga20}, 
zonal flows arising in magnetorotational instability (MRI) turbulence \citep[e.g.,][]{Johansen09, Krapp18}, 
and non-ideal MHD effects and magnetic disk winds \citep[e.g.,][]{Suriano17, Riols19,Cui21}. 

Magnetic fields have been observed in dense star-forming cores of molecular clouds that collapse to form stars and circumstellar disks \citep[e.g.,][]{Maury18, Galametz18, Gouellec20}. 
The magnetic field is widely believed to drive the disk evolution, particularly through a magnetized disk wind \citep[e.g.,][]{Bai13,Lesur22}, which removes angular momentum through a laminar torque. Two-dimensional (2D) non-ideal magnetohydrodynamic (MHD) numerical simulations have shown that axisymmetric disks may evolve into stable rings and gaps in wind-launching disks \citep[e.g.,][]{Suriano17, Suriano18}. More recent 2D non-ideal MHD simulations included detailed chemistry networks in computing the non-ideal MHD coefficients, finding that stable rings and gaps still form robustly, as long as the dimensionless Elsasser number $\Lambda$ (which measures the degree of field-neutral coupling) is of order unity or larger \citep[e.g.,][]{Nolan23}. The question that naturally arises is: are such axisymmetric rings and gaps stable in three dimensions (3D), particularly to the Rossby wave instability \citep[RWI,][]{Lovelace99, Li2000}?

The RWI occurs when the local Rossby wave is trapped in the non-self-gravitating disks. 
It may allow non-axisymmetric perturbations to grow into well-defined vortices \citep[e.g.,][]{Zaqarashvili21}. 
The RWI has been investigated in the 2D (vertically integrated) linear theories \citep[e.g.,][]{Lovelace99, Li2000, Umurhan10, Ono16} and numerical simulations \citep[e.g.,][]{Li01, Varnire06, Lyra08, Ono18, Cimerman23}. 
In the 3D disk (where the vertical structure is accounted for), the RWI has also been studied using both linear theories \citep[e.g.,][]{Lin2012, Lin13} and numerical simulations \citep[e.g.,][]{Lyra12, Richard13}, mostly with a prescribed initially axisymmetric substructure that is observationally motivated but not formed self-consistently.
 RWI has also been studied in magnetized disks, with toroidal magnetic fields \citep[e.g.,][]{Yu09}, large-scale poloidal magnetic fields \citep[e.g.,][]{Yu13}, and non-ideal MHD effects \citep[e.g.,][]{Can24}.
The spontaneous formation of rings and gaps in 2D (axisymmetric) non-ideal MHD simulations \citep[e.g.,][]{Suriano17, Suriano18, Nolan23} provides a rare opportunity to examine the RWI of axisymmetric structures produced through consistent dynamics using 3D simulations. Just as importantly, such 3D simulations also allow us to investigate the formation of disk substructures and the development of RWI in them simultaneously. In this paper, we seek to determine whether the non-linear development of RWI in the rings and gaps formed in the non-ideal MHD wind-launching disks can lead to the formation of vortices and, if so, how they affect the rings and gaps in the radial gas distribution. Vortices are important to investigate because of their potential for dust trapping \citep[see, e.g., a recent review by][]{Bae2023}, which may facilitate planetesimal formation.  

This paper is organized as follows. 
We present our simulation setup in \S \ref{sec:simultation}, including the governing equations. \S \ref{sec:non-ideal_coff} describes our chemistry model for computing the abundances of charged species, which are used to determine the non-ideal MHD coefficients. In \S \ref{sec:results}, we show results of both 2D and 3D simulations, focusing on RWI and its effects on the disk substructure formation. 
Our results are briefly discussed 
and summarized in \S \ref{sec:conclusion}.

%% file: Simulation.tex

We use Athena++ \citep{stone20} to solve the non-ideal MHD equations:
\begin{equation}
  \frac{\p \rho}{\p t} + \del \cdot (\rho \vel) = 0, \label{eq:continuity}
\end{equation}
\begin{equation}
  \frac{\p (\rho \vel)}{\p t} + \del \cdot (\rho \vel \vel + \ptot - \frac{\bb \bb }{4 \pi}) = - \rho \del \Phi, \label{eq:momentum}
\end{equation}
\begin{align}
  \frac{\p e_d}{\p t} + \del \cdot  \left[(e_d + \ptot)\vel - \frac{\bb (\bb \cdot \vel)}{4 \pi} + \frac{1}{c} \left( \eta_O \jj + \eta_{\rm AD} \jj_{\perp} \right)  \times \bb  \right] \notag \\
  = - \rho (\vel \cdot \del \Phi) - \Lambda_{\rm c}, \label{eq:energy}
\end{align}
and the induction equation
\begin{equation}
  \frac{\p \bb}{\p t}  = \del \times (\vel \times \bb) - \frac{4 \pi}{c} \del \times \left[ \eta_O \jj + \eta_{\rm AD} \jj_{\perp} \right] , \label{eq:induction}
\end{equation}
where $\rho$ and $\vel$ are gas mass density and velocity, $\bb$ is the magnetic field, $\ptot = P + B^2/(8\pi)$ is the total (thermal [$P$] and magnetic)  pressure, $\Phi = -GM/r$ is the gravitational potential of the central star, $e_d = \rho V^2/2 + {P}/(\gamma - 1) + B^2/(8 \pi)$ is the energy density, $\gamma$ is the adiabatic index, $\jj$ is the current density, $\jj_{\perp} = \bb \times (\jj \times \bb) / (B^2)$ is the component of $\jj$ perpendicular to the magnetic field, $\eta_O$ and $\eta_{\rm AD}$ are the Ohmic and ambipolar diffusivities, and $\Lambda_{\rm c}$ is the cooling term.

\subsection{Simulation domain} \label{subsection:simulation_domain}

\begin{figure}
    \centering
    \includegraphics[width=\linewidth]{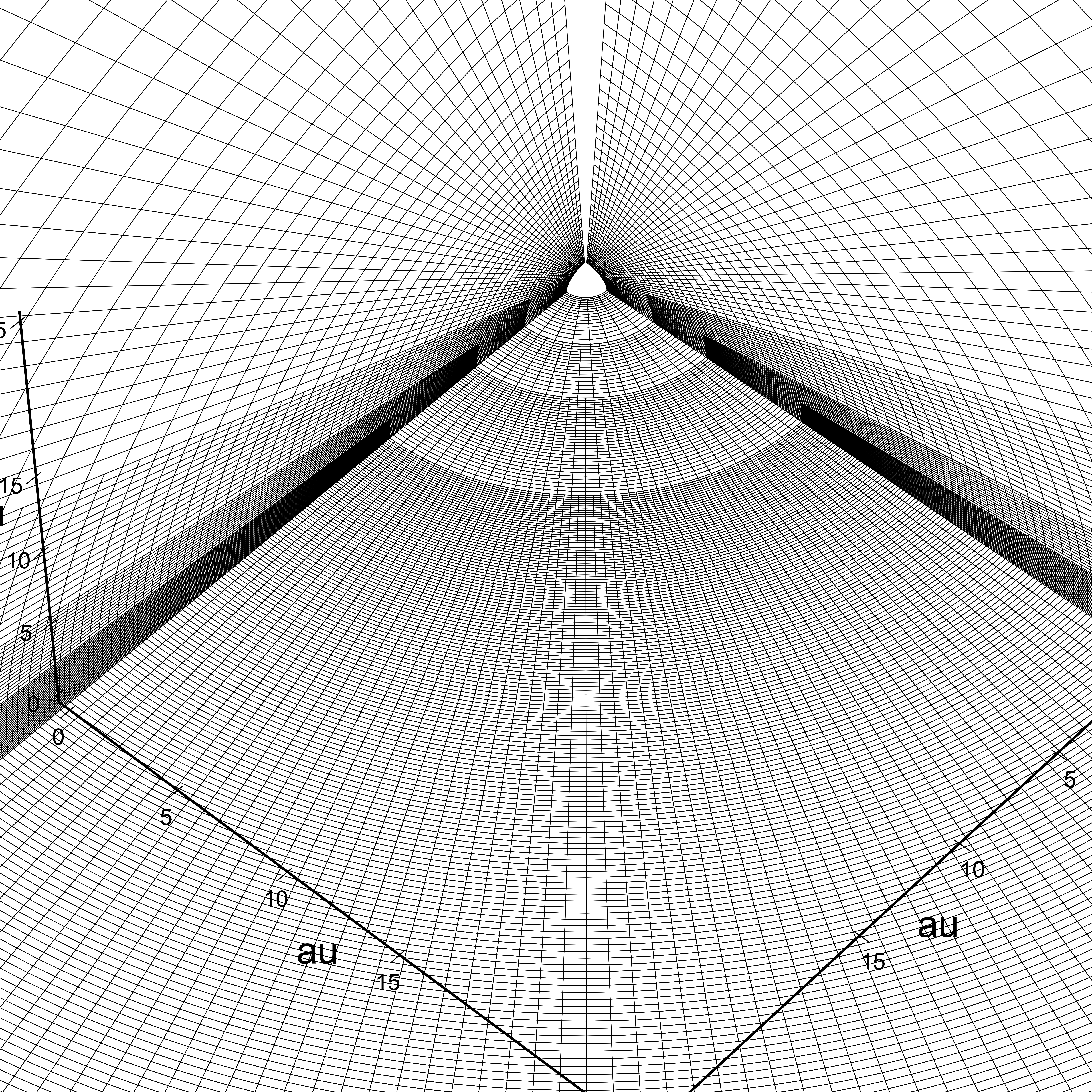}
    \caption{The inner part of the 3D grid, showing three levels of static mesh refinement.   
    }
    \label{fig_grid_structure}
\end{figure}

We use spherical coordinates ($r, \theta, \phi$) to perform the 2D (axisymmetric, with quantities independent of $\phi $) and 3D simulations. The simulation domain goes from 1 to 316 AU in the radial direction, 0.05 to $\pi$-0.05 in the polar direction, and, for 3D simulations, from 0 to $2\pi$ in the azimuthal direction. We use a logarithmic grid in the radial direction, with 80 base cells and a ratio of 1.07416 for the sizes of two adjacent cells. The grid is uniform in the polar direction, with 96 cells at the root level. For 3D simulations, we use 32 base cells in the azimuthal direction. Three levels of static mesh refinement (SMR) are used, with the grid size changing by a factor of 2 between adjacent levels, as indicated in Fig.~\ref{fig_grid_structure}. Specifically, the finest level extends from 10 to 100 au in the radial direction and about 2.5 scale heights ($\sim$ 0.13 radians) below to above the mid-plane. The disc scale height is resolved by about 12.6 grids. The second and third finest levels cover, respectively, 5.57 to 176 au and 3.14 to 316 au radially and a polar region within 0.25 radians and 0.5 radians of the mid-plane. 

\subsection{Boundary conditions} \label{subsection:boundary conditions}

We use modified outflow boundary conditions at both inner and outer radial boundaries. At the inner boundary, instead of copying gas density and pressure from the innermost active zone to ghost zones, we apply the same power law used in the gas initialization to extend gas density and pressure to ghost zones. For gas velocity, the azimuthal velocity follows a Keplerian curve, and the other two components are copied from the innermost active zone while preventing mass from outside from entering the simulation domain. Reflective conditions are used for the $\theta$ boundaries, and periodic boundary conditions are applied in the azimuthal direction.  

\subsection{Initial conditions} \label{initial_conditions_subsection}

We perform two types of simulations: (i) 2D simulations with an initial hydrostatic equilibrium state and a magnetic field generated from vector potential to ensure a divergence-free B field (i.e., $\nabla \cdot \bb = 0$) and (ii) 3D simulations that restart from 2D simulations either at relatively late times or near the beginning. 

We adopt the initial temperature and density profiles used in \cite{Hu22}. Specifically, we divide the simulation domain into a cold, dense disk and a hot, low-density corona, with a constant aspect ratio, $h/r=0.05$,  where $h$ is the disk scale height. We limit the cold, dense portion to two scale heights above (and below) the midplane, i.e., in the region $\pi/2 - \theta_0 <\theta < \pi/2 + \theta_0$, where $\theta_0=\arctan{(2h/r)}$. The gas density and temperature in the disk midplane both follow a power law with index $p=-1.5$ and $q=-1$, respectively:
\begin{eqnarray}
\rho(r,\pi/2) = \rho_0(r/r_0)^{p}  \\ \nonumber 
T(r,\pi/2) = T_0(r/r_0)^{q} 
\label{eq:profile}
\end{eqnarray}
where $r_0=1$~au is the radius of the inner boundary of the computational domain, and $\rho_0=2.667 \times 10^{-10} {\rm g/cm^{-3}}$ and $T_0=570$~K are the density and temperature at $r_0$. To ensure a smooth transition between the cold disk and hot corona, we adopt the following vertical profile for the temperature:  
\begin{equation}
T(r,\theta)=
\begin{cases}
T(r,\pi/2) & \text{if }  |\theta-\pi/2| < \theta_0 \\
T(r,\pi/2)\ exp[(|\theta-\pi/2|\\
\ -\theta_0)/\theta_0 \times\ln(160)]
  & \text{if } \theta_0 \leq |\theta-\pi/2| \leq 2\theta_0  \\
160\ T(r,\pi/2); & \text{if } |\theta-\pi/2| > 2\theta_0\\
\end{cases}\label{eq:T}
\end{equation}
We use a quick $\beta_{\rm cool}$ cooling scheme with a cooling timescale of only $10^{-10}$ of the local orbital period, so the temperature profile is effectively fixed over time.
The vertical density profile is generated based on hydrostatic equilibrium, i.e.,$v_r=v_\theta=0$. 

The initial magnetic field is computed from the magnetic vector potential used in \cite{Zanni07}:
\begin{align}
 & B_r (r, \theta) = \frac{1}{r^2 {\rm sin}\theta} \frac{\p A_\phi}{\p \theta}, \label{eq:Br} \\
 & B_\theta (r, \theta) = -\frac{1}{r\ {\rm sin}\theta} \frac{\p A_\phi}{\p r}, \label{eq:Btheta} \\
 & B_\phi = 0, \label{eq:Bphi}
\end{align}
with
\begin{align}
 & A_\phi (r, \theta) = \frac{4}{3}r_0^2 B_{\rm p,0} \left( \frac{r {\rm sin}\theta}{r_0}\right)^{\frac{3}{4}} \frac{1}{(1 + 2{\rm cot^2 \theta})^{5/8}}
\end{align}
where $B_{\rm p,0}$ sets the scale for the poloidal field strength. The magnetic field setup is the same as that of \cite{Bai17}, \cite{Suriano18}, and \cite{Hu22}. In all our simulations, $B_{\rm p,0}$ is set by plasma $\beta = 10^3$ in the mid-plane. 


We perform two types of 3D simulations to investigate (1) whether axisymmetric substructures formed in 2D simulations are stable to RWI and (2) how disc substructures and RWI develop simultaneously. For the first type, we restart in 3D a 2D simulation after it has evolved for a certain amount of time (typically $t_{\rm st}=2000$~years) when prominent rings and gaps have formed. For the second type, we restart from the beginning of the simulation ($t_{\rm st}=0$). Since the focus of the 3D simulations is on RWI, we add a random perturbation to the radial velocity up to 10$\%$ of the local sound speed at the time of restart to seed the instability. 

%% file: Chemistry.tex
To determine the non-ideal MHD coefficients in equation~(\ref{eq:induction}), the abundances of charged species must be computed first. We use a reduced chemical network similar to those of \cite{Umebayashi90} and \cite{Nishi91}. Specifically, we include the following elements: H, He, C, O, and Mg. Table. \ref{ch_table_abundance} in the Appendix lists their initial abundances. 

We include cosmic ray ionization, which dominates the ionization in the bulk of the disk. We adopt a total ionization rate \citep[including ionization of H$_2$ and He,][]{Umebayashi90} of $\xi = 10^{-17} {\rm s^{-1}}$. The rate equation for species $i$ of the fractional abundance, $x_i \equiv n_i/n_{\rm H}$, can be written as:
 \begin{align}
& \frac{1}{n_{\rm H}}\frac{{\rm d} x_i}{{\rm d}t} = \frac{1}{n_{\rm H}} \sum_j k_{\rm cr} x_j + \sum_{j,k} k_{jk} x_j x_k - x_i\sum_{m} k_{i,m} x_m, \label{eq:reaction_2}
 \end{align}
where $k_{\rm cr}$ is the cosmic-ray ionization rate, 
and $k_{jk}$ and $k_{i,m}$ are the formation and destruction rates by two-body reactions, respectively. 

We include three types of reactions in the paper: gas-phase, gas-grain, and grain-grain. Table \ref{ch_table_molecular_reaction} lists all gas-phase reaction rates used in the paper. The gas-phase reaction rates are adapted from UMIST database \cite{McElroy13}, with the rate coefficient represented by:
\begin{align}
 & k(T) = \alpha_{\rm c} \left(\frac{T}{300 \rm{K}} \right)^{\beta_{\rm c}} \rm{exp} \left(- \frac{\gamma_{\rm c}}{T} \right), \label{eq:Arrhenius_representation}
\end{align}
where $\alpha_{\rm c}$ is the constant factor, $\beta_{\rm c}$ is the power-index of the temperature dependence, and $\gamma_{\rm c}$ measures the energy barrier of the reaction.  


We adopt a power law of $n(a) \varpropto a ^{-3.5}$ for the grain size distribution, with the power index of the standard MRN distribution \citep{Mathis77}. There are 30 size bins logarithmically distributed from the minimum $a_{\rm min}$ to the maximum $a_{\rm max}$. A grain material density of 3~g/${\rm cm}^3$ is adopted. The dust-to-gas mass ratio is set to 0.01. The gas-grain and grain-grain reaction rates are calculated as average reaction rates following \cite{grassi19}: 
\begin{align}
 &\langle k_{c,g}(T) \rangle = \frac{ \int_{a_{\rm min}}^{a_{\rm max}}\varphi (a) k_{c,g}(a, T) {\rm d} a }{\int_{a_{\rm min}}^{a_{\rm max}}\varphi (a)  {\rm d} a}, \label{eq:averg_c_g_rates} \\
 &\langle k_{g,g}(T) \rangle = \frac{\int_{a_{\rm min}}^{a_{\rm max}} \int_{a_{\rm min}}^{a_{\rm max}}\varphi (a) k_{g,g}(a, a^{'}, T) \varphi (a^{'}) {\rm d} a {\rm d} a^{'}}{\int_{a_{\rm min}}^{a_{\rm max}} \int_{a_{\rm min}}^{a_{\rm max}}\varphi (a) \varphi (a^{'}) {\rm d} a {\rm d} a^{'}}  \label{eq:averg_g_g_rates}
\end{align}
where $k_{c,g}(a, T)$ denotes the rate coefficient for gas-grain reactions, 
and $k_{g,g}(a, a^{'}, T)$ is the grain-grain interaction rate coefficient. To simplify the symbol, we note $a_{\rm s} = a_i + a_j$ as the sum of the grain sizes for the grain-grain interaction. We assume a negligible radius for the gas-phase species compared to the grains, so $a_{\rm s} = a_j$ when we calculate $k_{c,g}(a, T)$. 
 Following \cite{Nishi91} and \cite{grassi19}, we consider up to two elemental charges sticking on the grains (e.g., $Z = 0, \pm 1, \pm 2$). Following \cite{Draine87,Zhao18}, we consider three cases for $k_{c,g}(a, T)$ and $k_{g,g}(a, a^{'}, T)$: 
 
 \begin{align}
& k^{0}_{i, j}(a_s, T) = \pi a_s^2 \left( \frac{8 k_{\rm B} T}{\pi m_{\rm r}}\right)^{1/2} \left[ 1 + \left( \frac{\pi e^2}{2 a_s k_{\rm B} T}\right)^{1/2}\right] S(T, Z_j), \label{eq:kij0} \\
& k^{-}_{i, j}(a_s, T) = \pi a_s^2 \left( \frac{8 k_{\rm B} T}{\pi m_{\rm r}}\right)^{1/2} \left( 1 - \frac{Z_j}{Z_i}\frac{e^2}{a_sk_{\rm B} T}\right) \notag \\ 
& \cdot \left[ 1 + \left( \frac{2}{\frac{a_s k_{\rm B} T}{e^2}- 2 \frac{Z_j}{Z_i}}\right)^{1/2}\right] S(T, Z_j), \label{eq:kij-} \\
& k^{+}_{i, j}(a_s, T) = \pi a_s^2 \left( \frac{8 k_{\rm B} T}{\pi m_{\rm r}}\right)^{1/2}  \notag \\
& \cdot \left[ 1 + \left( \frac{4a_s k_{\rm B} T}{e^2} + 3 \frac{Z_j}{Z_i} \right)^{-1/2} \right]^2 {\rm exp} \left( -\frac{\theta_{\nu} e^2}{a_s k_{\rm B}T}\right) S(T, Z_j), \label{eq:kij+} 
\end{align}
where $e$ is the elemental charge, $m_{\rm r} = m_im_j/(m_i + m_j)$ is the reduced mass, $S(T)$ is the sticking coefficient, and $\theta_\nu = Z_j^{3/2}/[Z_i(\sqrt{Z_i} + \sqrt{Z_j})]$. $k^{0}_{i, j}$ is for charged species colliding with neutral grains (e.g., $Z_iZ_j = 0$). $k^{-}_{i, j}$ is the rate for opposite charge reactions (e.g., $Z_iZ_j < 0$). $k^{+}_{i, j}$ is the reaction rate between species with the same electrical properties (e.g., $Z_iZ_j > 0$). If the collisions are between charged species and grains, then the index $i$ and $j$ represent the charged species and grains, respectively. The index $i$ and $j$ represent the lighter and heavier grains for grain-grain interactions. The sticking coefficients $S(T,Z)$ of electrons on grains are taken from \citet[][their Table 3]{grassi19}. Following \cite{Umebayashi90}, we set the sticking coefficients for non-electron charged species on grains to $S(T, Z) = 1$.

%% file: nonideal_MHD.tex
Once the charge abundances are known, the ambipolar, Hall, and Ohmic diffusivities can be computed using the standard procedure based on the parallel, Pedersen, and Hall conductivities \citep[e.g.,][]{Wardle07, Pinto08}:  
\begin{align}
    & \sigma_{\rm O} = \frac{ec}{B} \Sigma_j n_j|Z_j|\beta_{j, n}, \label{eq:parallel_conductivity} \\
    & \sigma_{\rm H} = -\frac{ec}{B} \Sigma_j \frac{n_j|Z_j|\beta_{j, n}^2}{1+\beta_{j, n}^2} \label{eq:hall_conductivity}, \\
    & \sigma_{\rm P} = \frac{ec}{B} \Sigma_j \frac{n_j|Z_j|\beta_{j, n}}{1+\beta_{j, n}^2} \label{eq:pederson_conductivity}
\end{align}
where
\begin{align}
 & \beta_{j, n} = \frac{eZ_jB}{m_jc} \frac{m_j + m_n}{\rho_n R_{j, n} (T)}  \label{eq:beta_j_n}
\end{align}
is the Hall parameter for collisions between the $j$-th charged species and neutral hydrogen, $n_j$ is the number density of the $j$-th species, $\rho_n$ and $m_n$ are the neutral (molecular) hydrogen mass density and mass, respectively, and $R_{j, n}(T)$ is the momentum exchange rate coefficient, which is discussed in detail in Appendix \ref{sec:momentum_transfer_rate_coefficients}. 

From the conductivities, we can compute the Ohmic, Hall, and ambipolar diffusivities as follows:
\begin{align}
 & \eta_{\rm O} = \frac{c^2}{4 \pi \sigma_{\rm O}}, \label{eq:Ohmic_diffusivity} \\
 & \eta_{\rm H} = \frac{c^2}{4 \pi \sigma_{\perp}} \frac{\sigma_{\rm H}}{\sigma_{\perp}}, \label{eq:Hall_diffusivity} \\
 & \eta_{\rm AD} = \frac{c^2}{4 \pi \sigma_{\perp}} \frac{\sigma_{\rm P}}{\sigma_{\perp}} - \eta_{\rm O}, \label{eq:ambipolar_diffusivity}
\end{align}
where $\sigma_{\perp} = \sqrt{\sigma_{\rm H}^2 + \sigma_{\rm P}^2} $. 

The dimensionless Elsasser numbers for these diffusivities are given by
$ \Lambda_{\rm O} = \frac{V_{\rm A0}}{\Omega \eta_{\rm O}}, \  \Lambda_{\rm H} = \frac{V_{\rm A0}}{\Omega \eta_{\rm H}}$ and $\Lambda_{\rm AD} = \frac{V_{\rm A0}}{\Omega \eta_{\rm AD}}$, where $V_{\rm A0} = B_0^2/(4 \pi \rho_0)$ is the Alfv\'en speed, with the lower subscript '0' denoting the local quantity at each location, and  $\Omega$ is the Keplerian angular velocity. 

To mimic the better magnetic coupling (and thus reduced diffusivity) due to higher ionization levels expected in the lower density regions near the disk surface and in the disk wind, we follow \cite{Suriano18} and multiply the magnetic diffusivities in Eq. (\ref{eq:Ohmic_diffusivity}) to (\ref{eq:ambipolar_diffusivity}) by the following $\theta$ dependence: 
\begin{align}
 & f(\theta) = 
\begin{cases} 
{\rm exp} \left(-\frac{{\rm cos}^2(\theta + \theta_0)}{2(h/r)^2} \right) & \mbox{if} \quad  \theta < \frac{\pi}{2} -  \theta_0 \\
1 & \mbox{if} \quad \frac{\pi}{2} - \theta_0 \leq \theta   \leq \frac{\pi}{2} + \theta_0 \\
{\rm exp} \left(-\frac{{\rm cos}^2(\theta - \theta_0)}{2(h/r)^2} \right) & \mbox{if} \quad \theta > \frac{\pi}{2} + \theta_0 
\end{cases}. \label{eq:theta_depend_non_ideal} 
\end{align}

%
%

%% file: Result.tex
\input{table_cases}

We will mainly consider models with the charge abundances computed from the chemical network discussed in \S~\ref{subsec:chemistry}. The abundances are shown in Appendix~\ref{sec:appendix_a_chemical} (Fig.~\ref{fig_ion_with_Shu}) for an MRN-type grain size distribution from $a_{\rm min}=0.5\ \mu$m to $a_{\rm max}=25\ \mu$m. These abundances are saved as a look-up table and referenced during the simulation to compute the non-ideal MHD coefficients in each computational cell. These models are labeled by the letter ``T " (for ``Table") in Table \ref{table_cases}, which lists all 2D and 3D models discussed in the paper. For comparison, we also consider a 2D model with the simple power-law prescription for the dominant molecular ion from \cite{Shu92} that was used in previous work along a similar line \citep{Suriano18, Hu22}. This model is denoted by the letter ``S" (for ``Shu") in Table \ref{table_cases}. The initial Elsasser numbers for these two model types are shown in Fig.~\ref{fig_non_ideal_comparison} of Appendix~\ref{sec:appendix_a_chemical} for reference.  

\subsection{2D Simulations}
\label{subsec:2D}

\begin{figure*}
    \centering
   \includegraphics[width=0.9\linewidth]{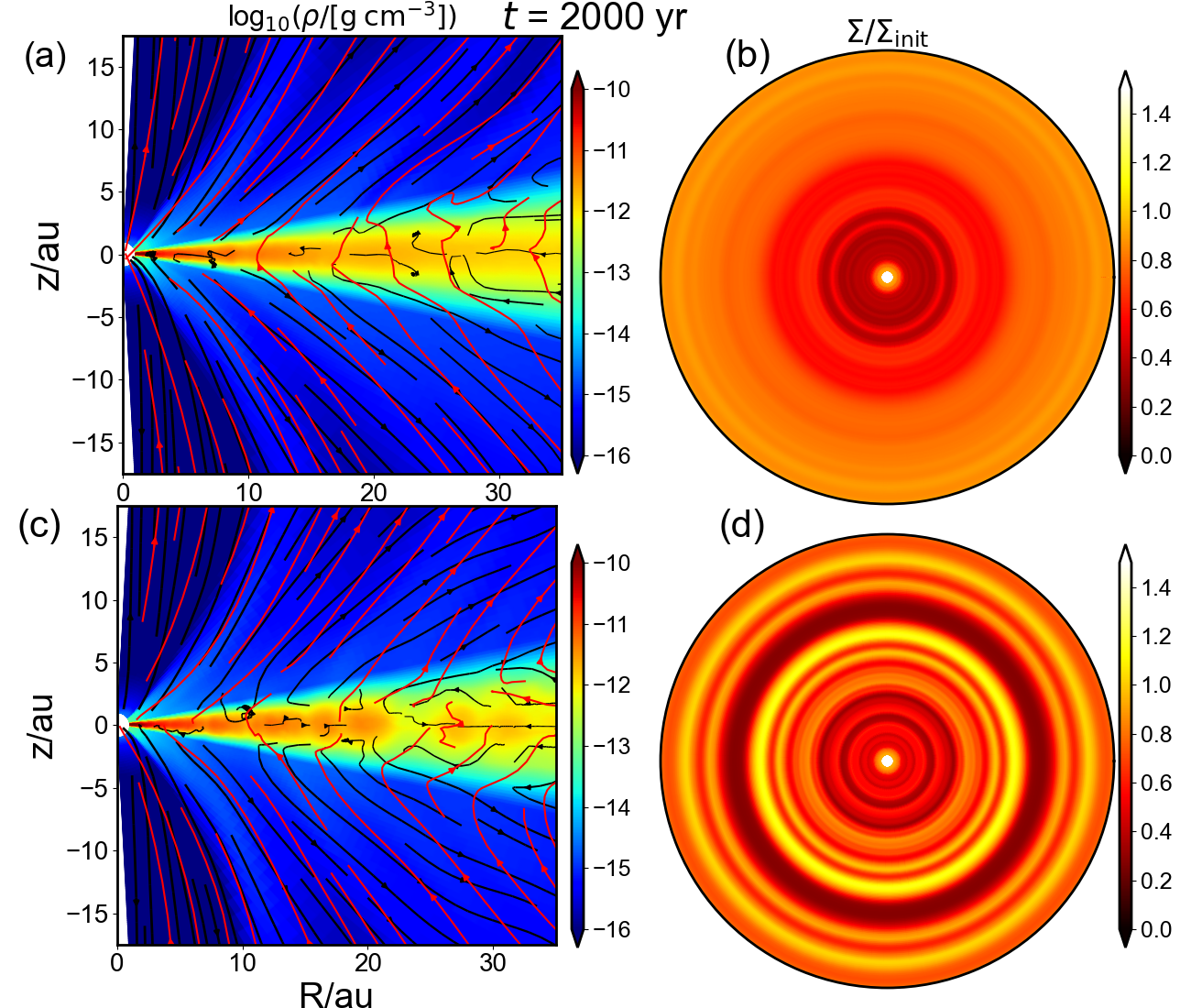}
    \caption{Rings and gaps in 2D Models S-2D (top panels) and T-2D (bottom panels) at a representative time $t=2000$~years in the inner disk (up to 35 au in radius). Plotted in panels (a) and (c) are the mass density distribution (color map), velocity streamlines (black lines with arrows), and magnetic field lines (red lines with arrows) on a meridional plane. A magnetized disk wind is clearly visible in both cases. Panels (b) and (d) display the surface density of the disk normalized to its initial value, highlighting the formation of prominent rings and gaps, especially in Model T-2D. 
    }
    \label{fig:2D_RingsGaps}
\end{figure*}

We start with a comparison of the 2D (axisymmetric) simulations using the Shu-type power-law prescription (Model S-2D) and the more realistic charge abundances from a look-up table computed from a chemical network including dust grains (Model T-2D). In Fig.~\ref{fig:2D_RingsGaps}, we plot a meridional (left panels) and face-on (right panels) view of the simulations at a representative time $t=2000$~years when stable rings and gaps are formed. We use cylindrical coordinates (R, $\phi$, z) in the meridional plots. 
The rings and gaps in Model T-2D are more prominent than those in Model S-2D because the gas in the former is better coupled to the magnetic field than in the latter in the bulk of the disk material (see Fig.~\ref{fig_non_ideal_comparison}). 
Our results add weight to the conclusion based on previous work in the literature \cite[e.g.,][]{Bai13, Lesur21}, especially that of \cite{Nolan23}, that ring and gap formation is a robust phenomenon in 2D that is not sensitive to the detailed treatment of the ambipolar diffusivity as long as the disk gas is reasonably well coupled to the magnetic field, with an ambipolar Elsasser number $\Lambda$ of order unity or larger. The question arises: Are the rings (and gaps) formed in 2D stable in 3D, particularly regarding Rayleigh instability and Rossby Wave Instability (RWI)?

\begin{figure}
    \centering
    \includegraphics[width=\linewidth]{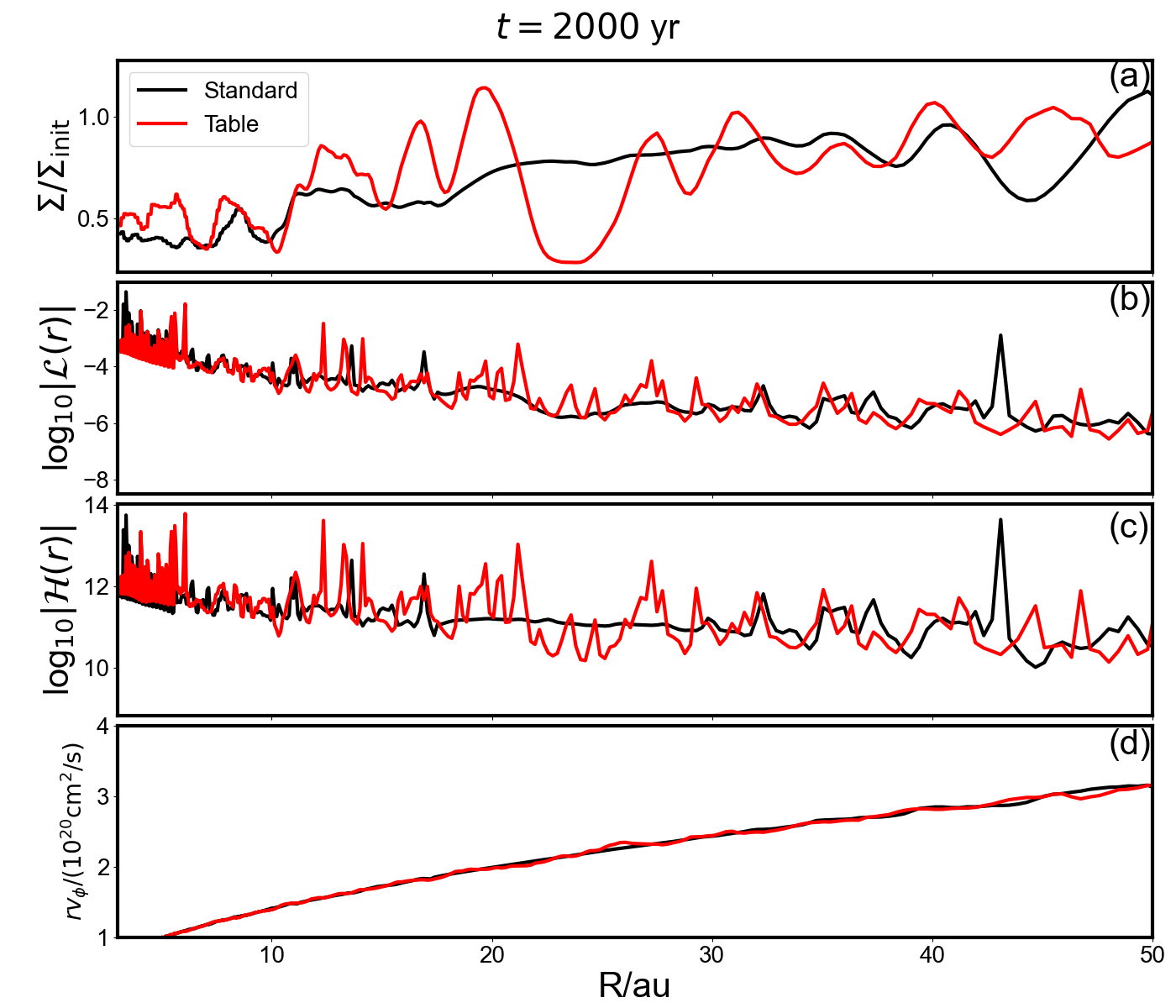}
    \caption{ Radial profiles of (a) the surface density of the disk normalized to its initial value, (b) the function $\mathcal{L}(r)$ (defined in equation [\ref{eq:LovelaceNumber}]) that controls the RWI for adiabatic flows, (c) the function $\mathcal{H}(r)$ (defined in equation [\ref{eq:isothermal}] that controls the RWI for locally isothermal flows, and  (d) the specific angular momentum on the midplane, for the disks with rings and gaps shown in Fig.~\ref{fig:2D_RingsGaps}. 
    Note the local maximum and the local minimum in the $\mathcal{L}(r)$ and $\mathcal{H}(r)$ profiles in panels (b) and (c) are the positions where the disk may be unstable to the RWI. Regions with specific angular momentum increasing with radius are stable to the Rayleigh instability for rotating flows.   
    }
    \label{fig:RWI_2D}
\end{figure}

One way to check the Rossby wave stability is to compute the quantity $\mathcal{L}(r)$ defined in \cite{Lovelace99} as a function of radius:
\begin{equation}
\mathcal{L}(r) \equiv \mathcal{F}(r){\rm S}^{2/\gamma}(r).
\label{eq:LovelaceNumber}
\end{equation}
The result is shown in Fig.~\ref{fig:RWI_2D}. 
Panel (a) shows the radial profiles of the column density of Model S-2D and Model T-2D at the time shown in Fig.~\ref{fig:2D_RingsGaps}. The column density contrast between the rings and gaps in Model T-2D is higher than in Model S-2D, consistent with Fig.~\ref{fig:2D_RingsGaps}. 
Panel (b) of Fig. \ref{fig:RWI_2D} shows the radial profiles of the quantity $\mathcal{L}(r)$
of Model S-2D and Model T-2D, where $\mathcal{F}(r) \approx \Sigma/(2 \omega_z)$, $\omega_z \equiv \hat{z} \cdot (\nabla \times \vel)$, and ${\rm S} \equiv P/\Sigma^{\gamma}$ with $P$ denoting the vertically integrated pressure and $\gamma$ is the adiabatic index. It is formally derived for an adiabatic flow for razor-thin disks, which differs from our case, where strong cooling keeps the flow nearly isothermal locally. In this case, the relevant quantity to evaluate is the inverse of a generalized vortensity, defined as \citep{Lin2012}: 
\begin{equation}
\mathcal{H}(r)\equiv \frac{2\Omega\Sigma c_s^2}{\kappa^2}=\frac{\Sigma c_s^2 r}{\partial (r v_\phi)/\partial r} = \frac{\Sigma c_s^2 }{\omega_z},
\label{eq:isothermal}
\end{equation}
where $\kappa$ is the local epicycle frequency. This quantity is plotted in panel (c). 
In addition, we plot in panel (d) the distribution of the specific angular momentum $r\ v_\phi$ on the disk midplane as a function of radius $r$. A few (limited) regions have specific angular momentum decreasing with radius; they are prone to Rayleigh instability, which may quickly develop to limit the amplitude of the pressure gradient-induced deviation from the background Keplerian rotation in 3D \cite[e.g.,][]{Ono18}. 

\cite{Lovelace99} and \cite{Lin2012} found that a local extremum of $\mathcal{L}(r)$ and $\mathcal{H}(r)$ is a necessary criterion for destabilizing the Rossby waves for adiabatic and locally isothermal flows, respectively. 
Panels (b) and (c) of Figure \ref{fig:RWI_2D} show local extrema are ubiquitous in our case. It indicates that the Rossby waves might grow throughout our simulation domain if there is an initial azimuthal variation to trigger them. 
However, our simulations do not strictly follow the situation envisioned in \cite{Lovelace99} or \cite{Lin2012} because of the inclusion of additional physical effects such as the magnetic field, non-ideal MHD effects, and disk winds. In what follows, we will determine whether the rings are indeed unstable to RWI through direct 3D simulations. 

\subsection{Stability of 2D Rings and Gaps in 3D} \label{subsec:3D_From_ReStart}

\begin{figure*}
    \centering
    \includegraphics[width=\linewidth]{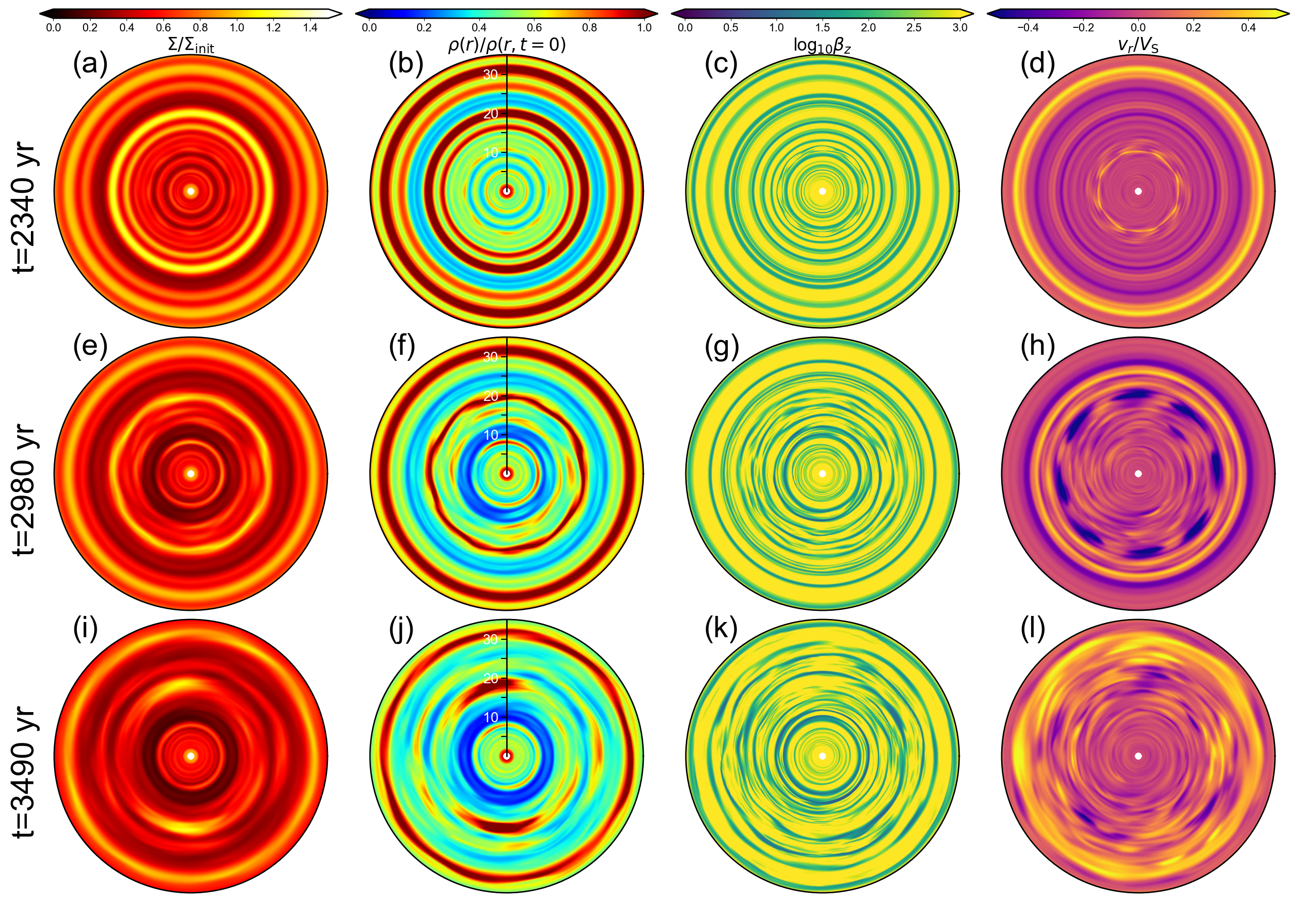}
    \caption{ 
    Evolution of 2D rings and gaps in 3D. Plotted in the first column are the surface density distribution normalized to its initial value at three representative times $t=2340$ (panel a), $2980$ (panel e), and $3490$~yr (panel i) of Model T-3D-2000 restarted from Model T-2D at $t=2000$~yr. Similarly, the second, third, and fourth columns show, respectively, the midplane mass density normalized to its initial value, the midplane plasma-$\beta_z$ based on the vertical field component $B_z$ alone, and the midplane radial velocity normalized by the sound speed. 
    An animated version of the figure can be found on the website: \url{https://virginia.box.com/s/b1wxk512g1xzm371thk9h7w4l1sfl8qk} 
    }
    \label{fig:RWI_Restart_1}
\end{figure*}

To check the stability of the disk substructures formed in the 2D simulations, we restart Model T-2D (which produced more prominent substructures than the other 2D model, Model S-2D) from the representative time shown in Fig.~\ref{fig:2D_RingsGaps}. We include a random perturbation to the radial component of the velocity up to $10\%$ of the local sound speed
to seed any potential instability that may grow.
The results of the restarted 3D model, T-3D-2000, are shown in Fig.~\ref{fig:RWI_Restart_1}. The figure plots the surface density distribution normalized to its initial value at the beginning of the 2D simulation ($t=0$), the midplane mass density normalized to its initial value, the midplane plasma-$\beta_z$ based on the vertical magnetic field component, and the midplane radial velocity normalized by the local sound speed at three representative times: $t=2340$ (top panels), $2980$ (middle), and $3490$~years (bottom panels). The link to an animated version of the figure can be found in its caption. 

Figure~\ref{fig:RWI_Restart_1} and its associated animation clearly show the development of non-axisymmetric structures, which follow the patterns expected for RWI; we will focus primarily on the rings since their azimuthal variations show up more clearly in the surface and mass density maps (RWI-induced structures in the gaps will be discussed towards the end of the next subsection). Specifically, the instability first develops in the inner disk and progressively moves to larger radii with longer local dynamical times. For a given ring of roughly constant radius, the azimuthal mode number $m$ of the dominant non-axisymmetric features decreases with time. For example, at the relatively early time of $t=2340$~yr (or 340~yr after the 3D restart), there are 8 well-defined over-dense clumps in the midplane mass density distribution that are regularly spaced on the ring at $\sim 11$~au (panel b), even though the clumps are less visible in the surface density at the same time (panel a). The $m=8$ mode is even more prominent in the distribution of the midplane plasma-$\beta_z$, which is much higher inside the over-dense clumps than between them (panel c). The same mode also shows up in the midplane radial velocity distribution, with the fastest radial outward motion occurring between the over-dense clumps and at a slightly smaller radius (panel d). We note that the time $t=2340$~yr corresponds to about 9 times the local orbital period at the unstable ring location, comparable to the saturation time for the fiducial model studied by \cite{Ono18}. 

Similar patterns are observed at the later time $t=2980$~yr but for a ring at a larger radius of $\sim 20$~au (compared to $\sim 11$~au discussed in the last paragraph). Here, the $m=8$ mode is the most prominent in the radial velocity distribution (panel h), with 8 regularly spaced regions of fast inflow at approximately half of the local sound speed. The fast-inflow regions are located slightly outside the visibly perturbed 20~au-radius ring, with the inflow apparently displacing the ring inward, producing a regularly spaced undulation in the radial direction, which is also evident in the surface density distribution (panel e). The undulation soon leads to the formation of 8 over-dense regions similar to those for the inner ring near 11~au at the earlier time $t=2340$~yr (see panel b), as can be seen from the animated version of the figure (see times around $t=3000$~yr). There is, therefore, little doubt that the 20~au-ring is RWI unstable, with a well-developed $m=8$ mode at $t=2980$~yr, corresponding to about 11 local orbital periods, similar to the 11~au ring at a comparable local dynamical time (normalized by the orbital period). The mode is less evident in the midplane plasma-$\beta_z$ distribution because it has more radial substructures than the density distribution, which makes the high plasma-$\beta_z$ ring near 20~au stand out less than the high mass density ring at a similar radius. 

The lack of one-to-one correspondence between the substructures in the plasma-$\beta_z$ and density distributions is illustrated by the faint ring inside the prominent gap around 10~au, which is barely visible in the density map (panel f) but stands out clearly as a high plasma-$\beta_z$ ring against the lower plasma-$\beta_z$ adjacent gaps (panel g). This low-density but high $\beta_z$ ring is the 11~au ring showing prominent $m=8$ azimuthal mode at the earlier time $t=2340$~yr (panel b). At this earlier time, at least seven well-defined rings were located inside the 20~au ring discussed at the later time $t=2980$~yr, with four outside the prominent 10~au gap and at least three inside. Visible azimuthal structures develop earlier in these (inner) rings than in the 20~au ring, with higher order modes at earlier times. By $t=2980$~yr, only relatively low-order azimuthal modes (with $m$ of a few) are clearly visible in the density map (panel f). In particular, two prominent arcs (corresponding to the $m=2$ azimuthal mode) develop in the ring immediately interior to the 10~au ring around $t=2700$~yr (see the animation of the density map), which has been smeared into a long arc with moderate azimuthal density variation by the time shown in panel (f). The trend for lower-order modes to dominate later is consistent with the expected evolution and saturation of RWI modes \cite[see, e.g.,][]{Ono18}.

The trend is vividly illustrated by the last frame of the simulation when the surface density and midplane density of the 20~au ring become dominated by two well-defined, relatively short, over-dense arcs (panels i and j) at $t=3490$~yr, corresponding to approximately 17 local orbital periods after the 3D restart. By this time, the azimuthal variations of the midplane density of the rings at larger radii remain dominated by higher $m$ modes. In particular, an $m=7$ mode starts to show up clearly in the outermost ring (at $\sim 32$~au) displayed at the time of panel (j), corresponding to approximately 8 local orbital periods, similar to the times (normalized by the local orbit period) when the $m=8$ mode becomes prominent for the inner 11~au (panel b) and 20~au (panel f) rings. 

It is unclear how long the low-order $m=2$ mode for the 20~au ring at the last frame would survive. A hint for the longer-term evolution of RWI-induced structures comes from the rings at relatively small radii, which evolved for longer local dynamical times than their larger-radius counterparts. For example, two prominent over-dense (relatively long) arcs develop in the 7.5~au ring around $2650$~yr (approximately 31 local orbital periods after the 3D restart), and last until $2760$~yr for $\sim 110$~yr or $\sim 5$ local orbital periods (see the animated version of the midplane density). A single ($m=1$) over-dense arc dominates the appearance of the ring at some of the later times (e.g., $t=3300$~yr), in agreement with the expectations based on the simpler hydro case \cite[e.g.,][]{Ono18}. However, this is not the case at other times, e.g., at the last frame shown in panel (j), where at least two, possibly three, arcs with moderate density enhancements exist. This deviation from the expectations is unsurprising since our simulated disk is more complex than the simplest hydro case (with, e.g., non-ideal MHD and disk wind). In particular, the rings and gaps in our simulation are highly dynamic structures that can gain or lose mass and angular momentum through spatially dependent mass accretion and (magnetized) outflow. 

\begin{figure*}
    \centering
    \includegraphics[width=1.0\linewidth]{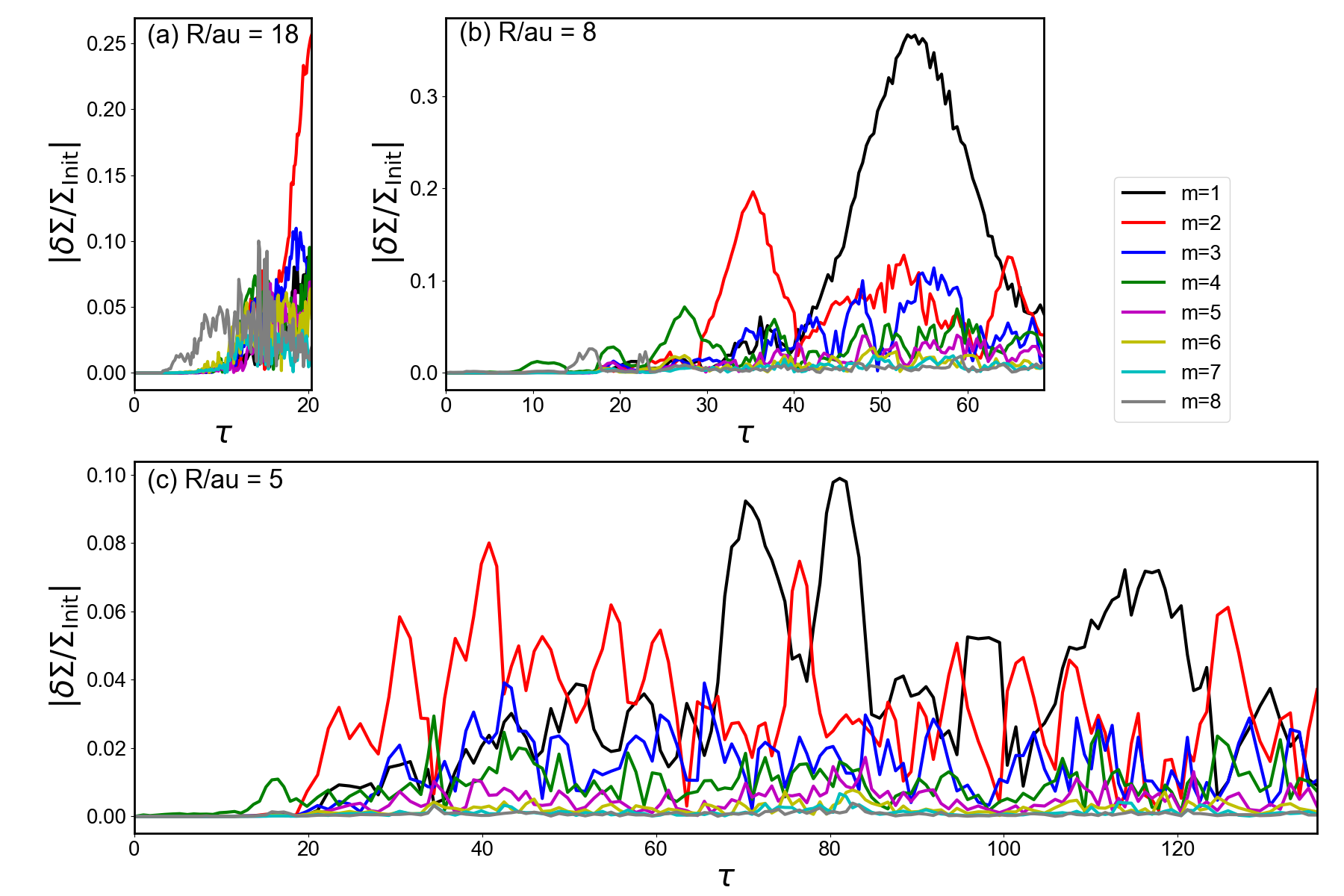}
    \caption{The time evolution of the mode amplitudes of the azimuthal surface density variation for three representative rings at radii of $R\sim 18$ (panel a), 8 (b), and 5~au (c) of Model T-3D-2000. The time is normalized by their local period. Panel (a) shows the $m=8$ mode first dominates the 18~au ring before being overtaken by the $m=2$ mode. Panels (b) and (c) illustrate a similar trend: the higher-order modes develop first before being overtaken by low-order modes, as expected for RWI modes. 
    }
    \label{fig:LateTimeModes}
\end{figure*}

To quantify the late-time evolution of different azimuthal modes, we plot in Fig.~\ref{fig:LateTimeModes} the mode amplitudes of the azimuthal surface density variations of three representative rings at approximately 5, 8, and 18~au as a function of time normalized by their local orbital period. The low-order $m=2$ mode (see the red curve) starts to dominate at a time between $\sim 15$ and $\sim 30$ local orbital periods; its further evolution is truncated by the termination of the simulation for the 18~au ring (panel a). For the 8~au ring, the $m=2$ mode decays after $\sim 10$ local orbital periods and is replaced by the even lower-order $m=1$ mode with a higher peak amplitude (panel b). The $m=1$ mode decays after $\sim 30$ orbits. Its longer-term evolution is unclear because of the simulation time limit, but a hint is offered by the 5~au ring, which reached $\sim 130$ orbits. In this case, the $m=1$ mode is more important than the $m=2$ mode most of the time after $\sim 65$ orbits, but the $m=2$ remains significant and occasionally surpasses the $m=1$ mode in amplitude. A caveat is that the 5~au ring is on a coarser grid than the other two rings and is resolved azimuthally by only 64 cells (see the statically refined grid in Fig. \ref{fig_grid_structure}). The relatively low resolution may have affected the growth and saturation of the modes. Nevertheless, it is clear that, as expected for RWI, low-order modes dominate the azimuthal variation of the surface density at late times, with relatively small amplitudes. An implication is that the non-linear development of RWI modifies but does not destroy the rings formed in our non-ideal MHD simulations.

\begin{figure*}
    \centering
    \includegraphics[width=0.9\linewidth]{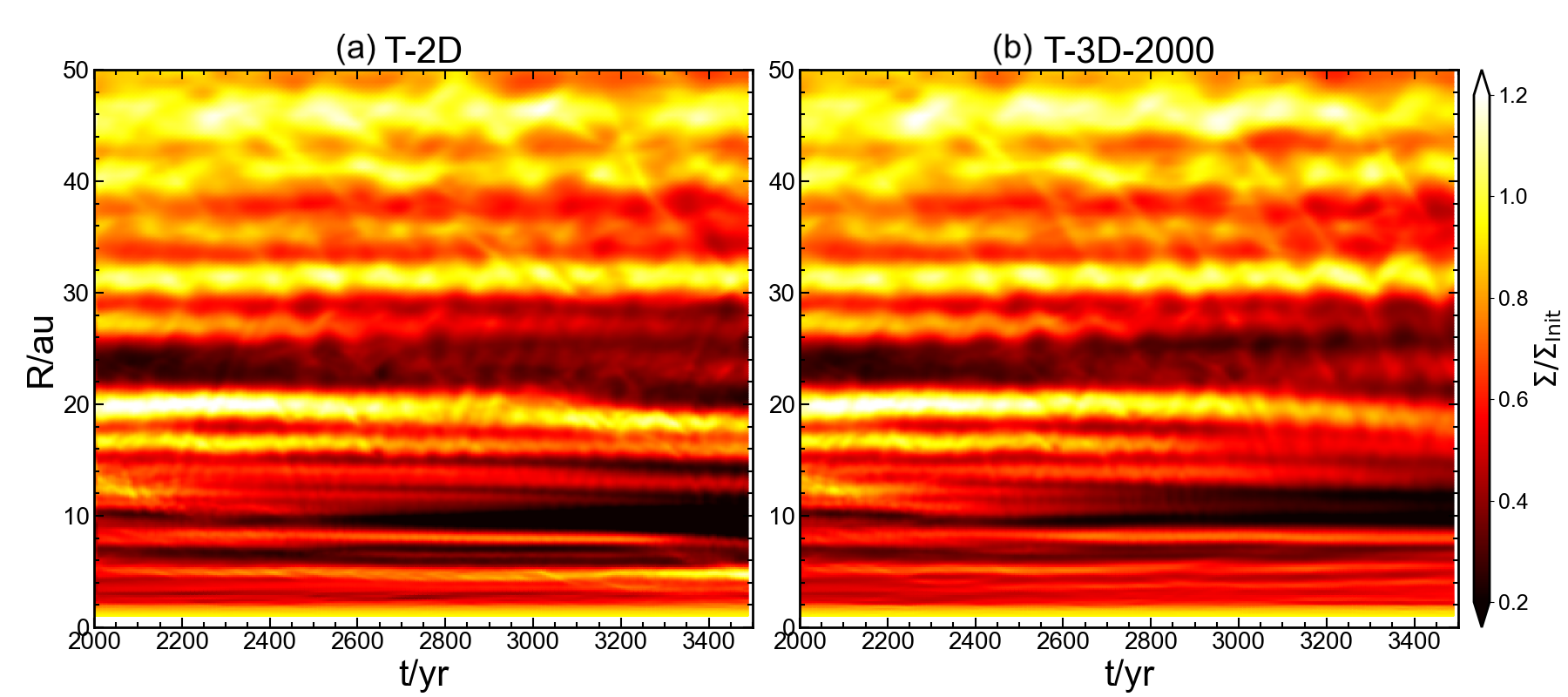}
    \caption{Time evolution of (a) the surface density distribution of Model T-2D and (b) the azimuthally averaged surface density distribution of Model T-3D-2000 as a function of radius after the 3D restart at  = 2000 yr. Note the prominent rings and gaps survive in 3D despite RWI. 
    }
    \label{fig:RingGapSurvival}
\end{figure*}

The effects of the RWI modes on rings are further illustrated in panel (j) of Fig.~\ref{fig:RWI_Restart_1}, which shows that the 8~au ring remains distinct from its neighboring gaps 
(i.e., with a significant axisymmetric $m=0$ mode) even after a relatively large number ($\sim 72$) of local orbit periods, indicating that nonlinear development of RWI modifies rather than destroys the ring. The same is true for rings at smaller radii, such as the one around 5~au, which retains its ring-shaped structure after $\sim 133$ local orbital periods. To demonstrate the survival of the rings and gaps more clearly, we plot in Fig.~\ref{fig:RingGapSurvival} the azimuthally averaged surface density distribution of Model T-3D-2000 as a function of radius after its restart at $t=2000$~yr and compare it to its 2D counterpart Model T-2D. There are some differences between the two models at late times, including a more prominent ring at $\sim 5$~au in the 2D case and a less empty gap at $\sim 10$~au in the 3D case. Nevertheless, the azimuthally averaged radial substructures are broadly similar in the two cases, supporting the notion that rings and gaps survive in 3D despite RWI. 

%
%

\subsection{RWI Vortex Structure}
\label{sec:vortex}

\begin{figure*}
    \centering
    \includegraphics[width=\linewidth]{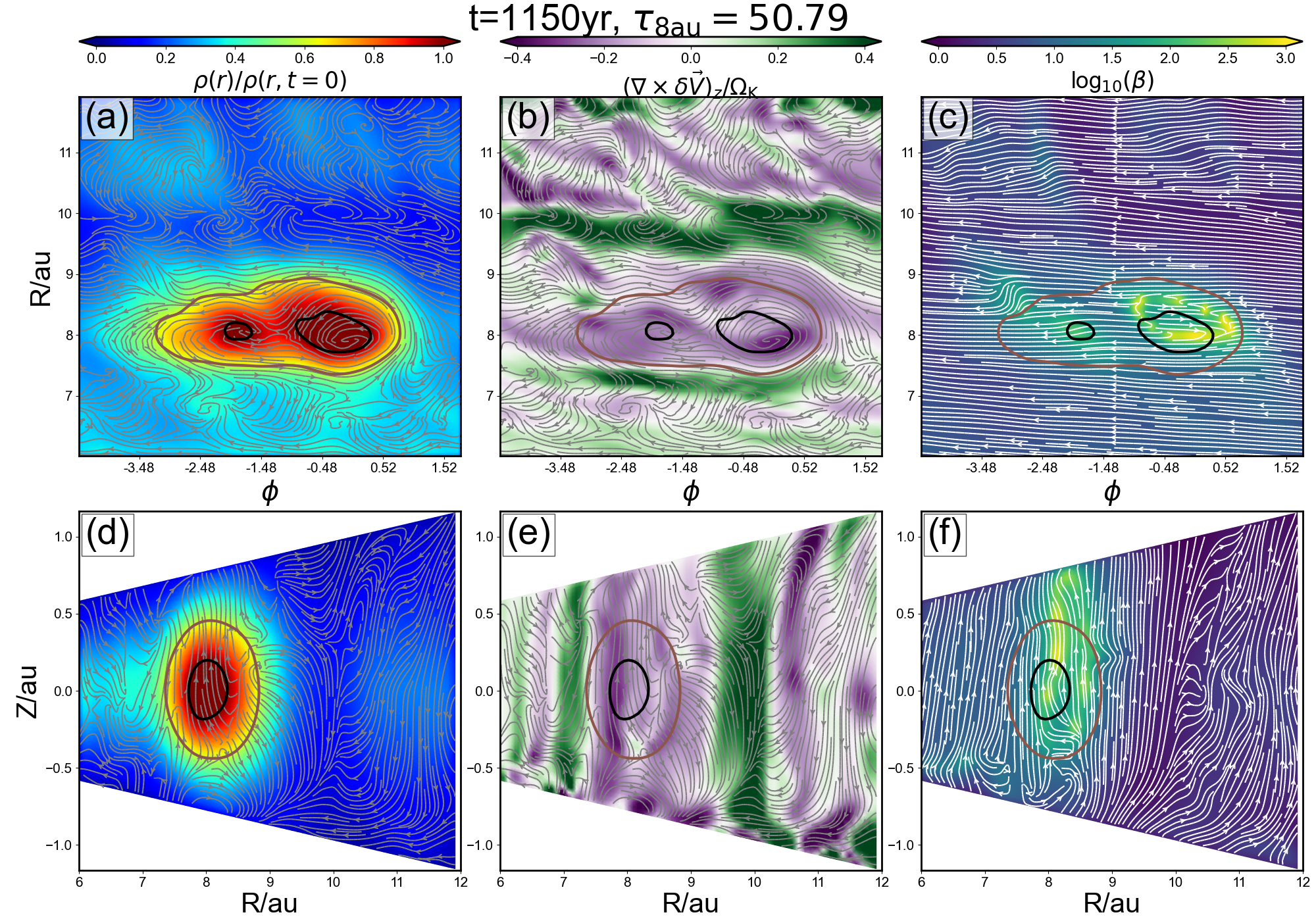}
    \caption{Structure of the dominant vortex in the 8~au-ring at $\tau \sim 50$. Panels (a)-(c) show the mass density normalized to its initial value, the $z-$component vorticity normalized to its local Keplerian angular speed, and the logarithmic scale of the plasma-$\beta$ on the midplane, respectively. Panels (a) and (b) include the velocity streamlines (gray lines) and the panel (c) the magnetic fields (white lines). Panels (d)-(f) are the same as (a)-(c) but on a meridional plane passing through the density peak of the 8 au ring. The brown and black contour lines are 0.6 and 1.0 contours of the mass density normalized to its initial value. An animated version of the figure can be found at \url{https://virginia.box.com/s/499xq1qt2j9c9vgxzljlb19x610tfnqe}
    }
    \label{fig:VortexStructure_1}
\end{figure*}

\begin{figure}
    \centering
    \includegraphics[width=\linewidth]{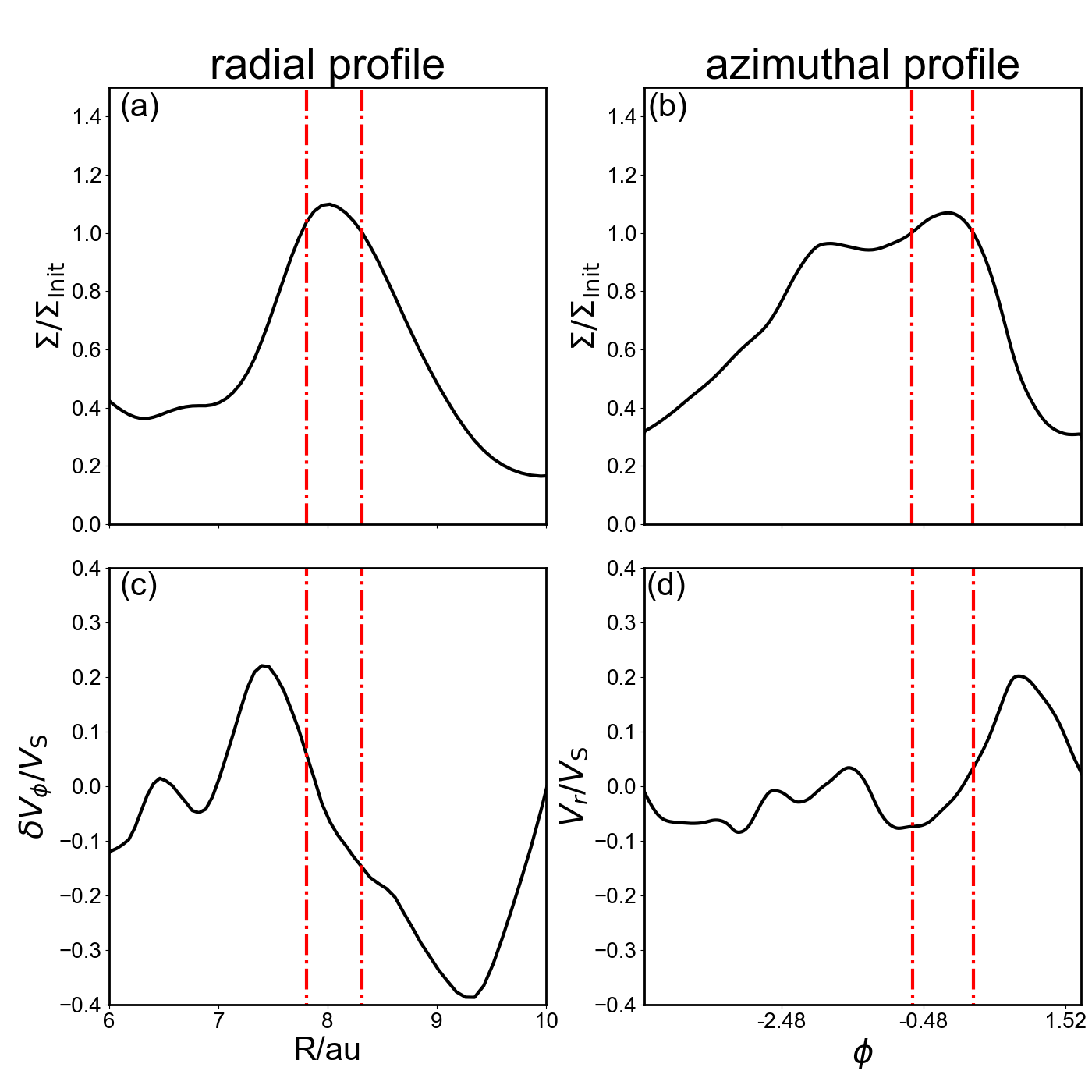}
    \caption{Detailed structure of the dominant RWI vortex in the 8~au ring of the Model T-3D-2000 at $\tau \sim 50$. Panels (a) and (c) show the radial profiles of the surface density normalized to its initial value and azimuthal velocity deviation from the local Keplerian value normalized to local sound speed, respectively, on a meridional plane passing through the density peak at 8 au. Panels (b) and (d) show the azimuthal profiles of the surface density normalized to its initial value and radial velocity normalized to the local sound speed of the same vortex. The red dash-dotted lines mark the locations where the mass density normalized to its initial value is 1.0 (corresponding to the black density contour in Fig.~\ref{fig:VortexStructure_1}).
    }
    \label{fig:VortexStructureCut}
\end{figure}

In this subsection, we analyze the structure of the vortices for the 8~au ring at the time $t=3150$~yr (or 1150~yr after the 3D restart), when the $m=1$ azimuthal mode reaches its peak amplitude (see Fig.~\ref{fig:LateTimeModes}b). We will concentrate on the broad features of the vortices since their detailed structures may not be adequately resolved in our 3D global non-ideal MHD simulations. At this time, the $m=2$ mode has a significant amplitude as well, leading to a secondary density peak next to the primary one, as shown in Fig.~\ref{fig:VortexStructure_1}a and Fig.~\ref{fig:VortexStructureCut}b (the one with a smaller value of the azimuthal angle $\phi$). As expected, these vortices are anti-cyclonic, with the flow streamlines on the mid-plane circling the local density maxima in a direction opposite to the Keplerian rotation (see Fig.~\ref{fig:VortexStructure_1}a). 

The corresponding negative axial vorticity is shown in panel (b) of Fig.~\ref{fig:VortexStructure_1}. Note that the spatial distributions of the axial vorticity in and around the vortices are patchier compared to 3D hydro simulations \cite[e.g., Fig.~8 of ][]{Richard13}, likely because the magnetic fields included in our simulation make the disk more dynamically active and can produce spatially variable vorticity even in the absence of RWI-induced vortices. Nevertheless, regions of high negative vorticity appear relatively coherent in the vertical direction, as shown in their $R-z$ distribution (panel e), similar to the columnar structure seen in the hydro case \cite[e.g., Fig.~8 of ][]{Richard13}.

The streamlines in panels (d) and (e) show that the flow on the poloidal ($R-z$) plane has a downdraft in the outer part of the vortex (with larger radii than that of the density peak) and an updraft in the inner part (near the density peak), with some hint of a clockwise vortex in between. This pattern is reminiscent of the one shown in the lower-left panel of Fig. 13 of \cite{Meheut10} (in the region between their $\sim 2.8$ and $3.0\ r_i$), who studied the 3D structures of hydro RWI vortices. However, the vortex is barely visible and much less prominent than the hydro case. There is a prominent counter-clockwise vortex outside the clockwise one in the hydro case, but not in our case. The difference may not be surprising since our wind-launching non-ideal MHD disk has strong meridional circulation motions, particularly in dense rings (e.g., Fig.~8 of \citealt{Hu22}; see also \citealt{Riols19,Cui21} and Fig.~\ref{fig:Vortex_T3D-0}d below), which can, in principle, change the dynamics and structure of the RWI vortices. 

Specifically, magnetic fields may affect the motions in and around the vortices. For example, the field lines in the $R-\phi$ plane are mostly toroidal (i.e., along $\phi$-direction; panel c), which may resist the field bending by the cross-field motion associated with the anti-cyclonic rotation around the density peak of the vortex (panel a). However, the magnetic fields inside the over-dense vortex are relatively weak, corresponding to a typical plasma-$\beta$ of order $10^2$ (panel c), so their direct resistance to the vortex motion may be relatively modest. Nevertheless, the magnetic field and its associated outflow dominate the angular momentum transport and generate strong poloidal motions and non-Keplerian rotation throughout the disk, which could affect the RWI vortices indirectly; indeed, they give rise to the RWI-unstable substructures in the first place. However, such indirect effects are difficult to quantify. 

An interesting feature of the plasma-$\beta$ distribution in the $R-\phi$ plane (panel c) is a ring of high $\beta$ values offset from the density peak, where the field lines change directions abruptly. Panel (f) shows that the feature extends vertically outside the high-density part of the vortex outlined by the isodensity contours, indicating that it is primarily a feature of weak local magnetic field rather than high density. Specifically, it is primarily associated with regions where the magnetic field reversal happens, particularly in the 
$\phi$-direction (i.e., where $B_\phi$ approaches zero). The field reversal in or near dense vortices is unsurprising because the gas motions there are strong enough to tangle the field lines, creating kinks where the magnetic field components (such as $B_\phi$ or $B_r$) drop to zero. Field tangling is harder to achieve in the more strongly magnetized lower-density regions (e.g., gaps), where the field lines are straighter. 

To analyze the vortices more quantitatively, we plot in Fig.~\ref{fig:VortexStructureCut} the radial and azimuthal profiles of the normalized surface density across the density peak of the primary vortex (panels a and b) and the radial profile of the deviation of the azimuthal velocity from the local Keplerian value (panel c) and the azimuthal profile of the radial velocity (panel d) normalized by the local sound speed. The radial density profile is asymmetric because the gap outside the vortex-forming ring is deeper than that inside (see Fig.~\ref{fig:VortexStructure_1} and its animated version). The super-Keplerian rotation radially interior to the density peak and sub-Keplerian rotation exterior to the peak are consistent with expectations.  
As expected from the counterclockwise circulation pattern of the anti-cyclonic vortex around the density peak in the $R-\phi$ plane (see Fig.~\ref{fig:VortexStructure_1}a), the radial velocity is positive on the larger azimuthal angle side of the density peak, reaching a maximum of $\sim 0.2$ times the local sound speed. It is negative on the other side, decreasing to a minimum of only $\sim -0.1$ times the local sound speed; the asymmetry is caused by the secondary vortex nearby. 

\begin{figure}
    \centering
    \includegraphics[width=\linewidth]{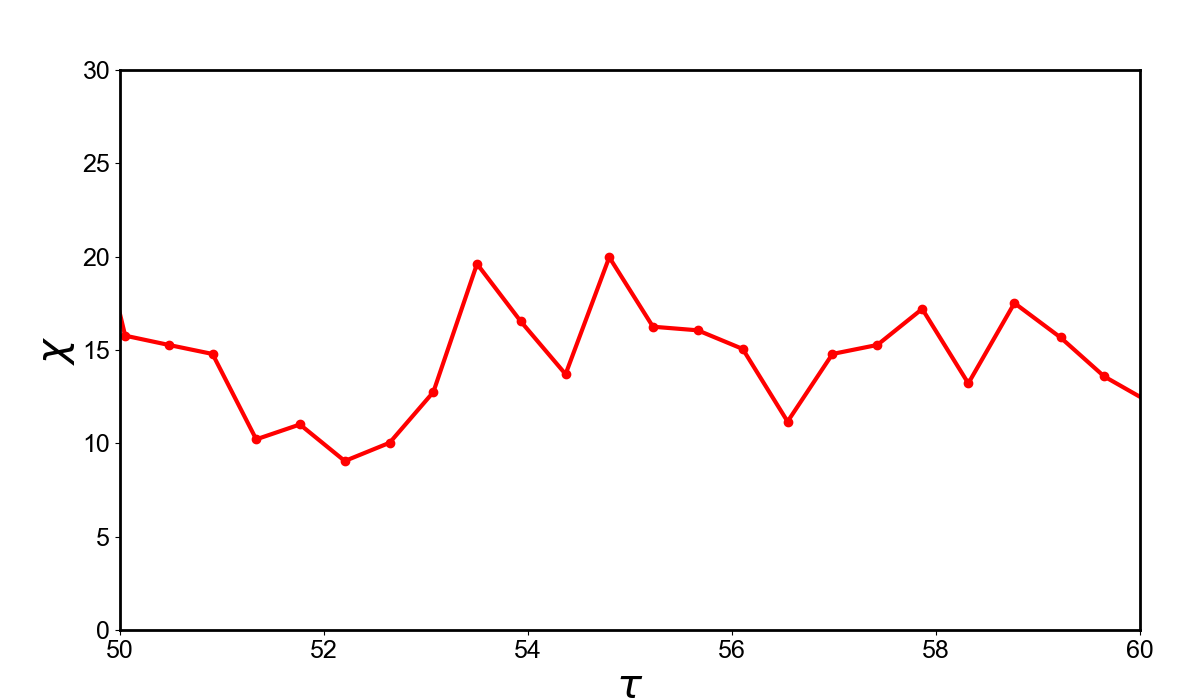}
    \caption{The aspect ratio for the dominant vortex at 8 au in Model T-3D-2000 from $\tau = 50$ to $60$ (the time is normalized by the local orbital period). 
    }
    \label{fig:Vortex_aspect_ratio}
\end{figure}

 An important parameter to characterize a vortex is its aspect ratio $\chi \equiv a/b$, where $a$ and $b$ are the major and minor axes of close streamlines \cite[e.g.,][]{Ono18}. Because the velocity field in our non-ideal MHD simulation is much more disordered than the hydro simulations (see the velocity streamline in Fig. \ref{fig:VortexStructure_1}), we choose to approximate the shape of the vortex using the iso-density contour of $\rho(r,\phi)/\rho(r, t=0)=1$ (i.e., the black lines for the dominant vortex in Fig. \ref{fig:VortexStructure_1}a). The resulting aspect ratio $\chi$ for the dominant vortex near the 8~au radius is shown in Fig.~\ref{fig:Vortex_aspect_ratio} as a function of time during the period when the $m=1$ mode dominates 
($50<\tau<60$; see panel b of Fig. \ref{fig:LateTimeModes}). The aspect ratio is of order 10 or larger for the entire duration, indicating that the vortex is highly elongated, consistent with visual inspection of Fig.~\ref{fig:VortexStructure_1} and especially Fig.~\ref{fig:RWI_Restart_1} and their associated animations. \cite{Lesur09} pointed out that vortices in disks tend to be unstable to ellipsoidal instability in 3D, although the maximum growth rate obtained from linear analysis decreases rapidly with increasing aspect ratio $\chi$ in a stratified disk (see the red curve in their Fig.~6), making it increasingly difficult to detect in simulations. \cite{Richard13} found that highly elongated vortices with $\chi \sim 6$ or larger can survive in 3D compressible hydro simulations. This hydro result contrasts with our non-ideal MHD simulation, where the amplitude of the dominant $m=1$ mode at 8~au steadily decreases after peaking around $\tau\approx 55$ despite its large aspect ratio. 

There are several possibilities for the above difference. Firstly, the magnetic field in our simulation may weaken the vortex. For example, 
the circulating flow in the vortex must move across the predominantly toroidal magnetic field in the midplane (contrasting the flow streamlines and field lines in, e.g., Fig.~\ref{fig:VortexStructure_1}a and c, respectively) and is expected to be resisted by magnetic tension.
 \cite{Lyra11} found that the elliptic instability is quite destructive in magnetized disks; perhaps this explains why even elongated vortices decay in our simulations.
Secondly, the vortex may be weakened by the magnetized disk wind in our simulation. Through mass and angular momentum removal, the wind induces strong meridional circulation that may be incompatible with the columnar structure preferred by the vortex. In particular, significant vertical motions are present in our simulation (see, e.g., Fig.~\ref{fig:VortexStructure_1}d), which are different from those found in 3D hydro simulations \cite[e.g.,][]{Meheut10,Meheut12,Richard13}. Thirdly, our wind-launching non-ideal MHD disk is more dynamically active than the relatively quiescent disk envisioned in most of the previous 3D hydro RWI simulations, with an effective $\alpha$ parameter of order 0.01 to 0.1 estimated from the Reynolds and Maxwell stresses, which can negatively impact the vortex's survival \cite[e.g.,][]{Lin13}. 

Interestingly, many {\em cyclonic} vortices form in our restarted 3D simulation in addition to the anti-cyclonic vortices discussed above. They are formed in low-density gaps and evolve broadly like the anti-cyclonic vortices formed in dense rings: higher-order azimuthal modes dominate at earlier times and lower-order modes at later times. For example, eight cyclonic vortices (i.e., the $m=8$ mode) are clearly visible in the $\sim$10-au gap from the streamlines plotted on the $R-\phi$ plane at a relatively early time $t=450$~years after the restart (corresponding to about 14 local orbital periods), as shown in the animated version of Fig.~\ref{fig:VortexStructure_1}. They merge into fewer numbers at later times. In particular, at the time of $t=1150$~years shown in Fig.~\ref{fig:VortexStructure_1}, there are only two readily identifiable cyclonic vortices with (clockwise) closed streamlines in the 10-au gap, as can be seen from panel (a) and particularly (b), which also shows that the cyclonic vortices have the largest (positive) vertical component of the vorticity $\omega_z$. Besides the sign of $\omega_z$, these vortices differ from the anti-cyclonic ones (with a negative $\omega_z$) in one major aspect: they are more strongly magnetized, with a plasma-$\beta$ typically of order unity, as can be seen from panels (c) and (f).  
Nevertheless, despite the strong magnetization in the gap, cyclonic vortices appear to be able to develop and survive until the end of the simulation at $t=1490$~years after the 3D restart, corresponding to $\sim$47 local orbital periods at 10~au. 
\vskip 0.5cm

Since the perfectly axisymmetric rings and gaps formed in 2D simulations are shown to be unstable to RWI, such idealized structures are unlikely to be produced in nature without the assumption of axisymmetry. We are thus motivated to start the 3D simulations from the very beginning to determine how the RWI interacts with the formation of the rings and gaps in the first place. 

\subsection{3D Simulation from the Beginning}
\label{subsec:3D_From_Start}

\begin{figure*}
    \centering
    \includegraphics[width=\linewidth]{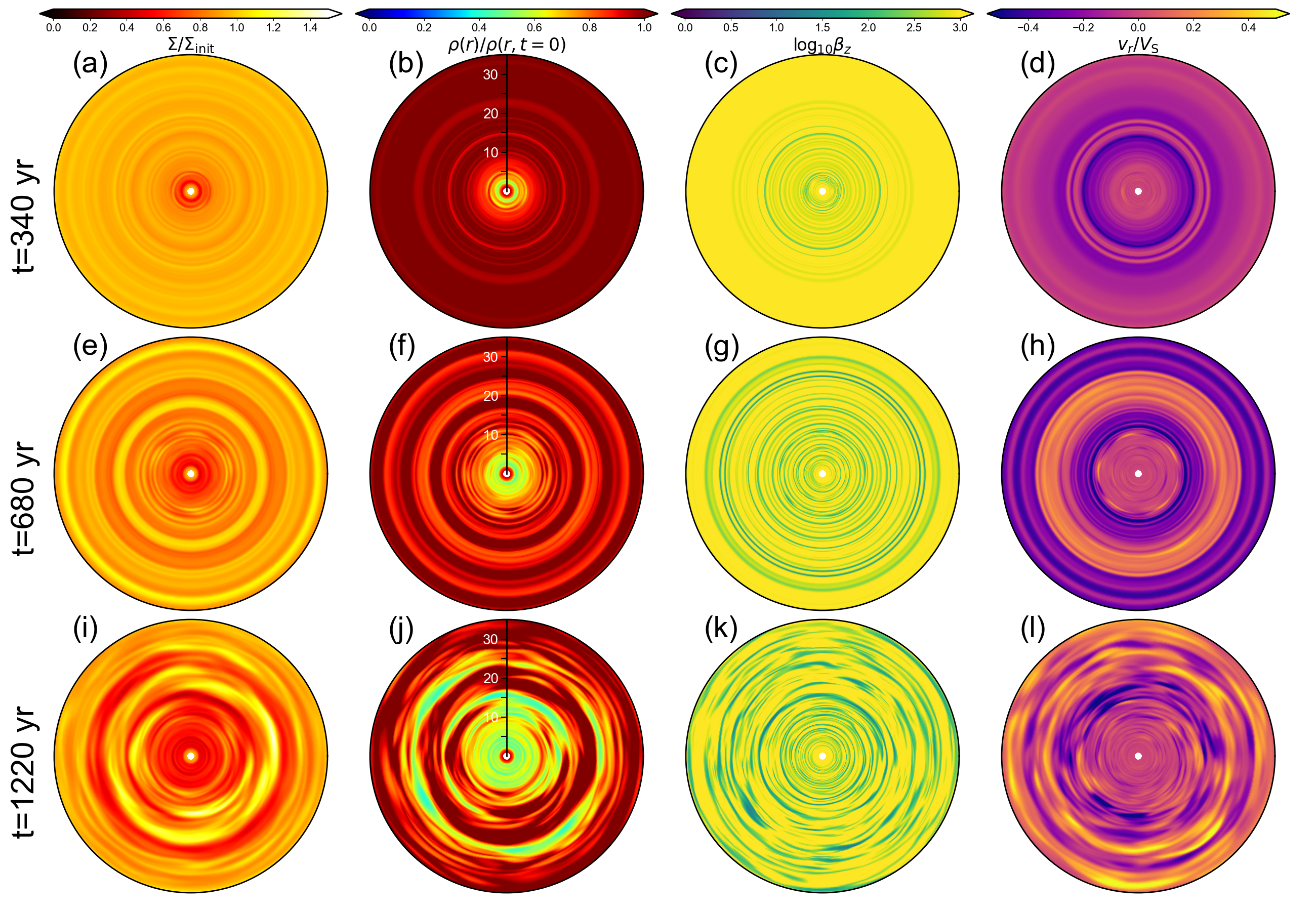}
    \caption{Simultaneous growth of rings and gaps and RWI in 3D in Model T3D-0. Plotted in the first column are the surface density distribution normalized to its initial value at three representative times $t=340$ (panel a), $680$ (panel e), and $1220$~yr (panel i). Similarly, the second, third, and fourth columns show, respectively, the midplane mass density normalized to its initial value, the midplane plasma-$\beta_z$ based on the vertical field component $B_z$ alone, and the midplane radial velocity normalized by the sound speed. 
    An animated version of the figure can be found on the website: \url{https://virginia.box.com/s/shw9ux7dc0klb3bduj13iqvq3hj45rjq} 
    }
    \label{fig:RWI_Restart_2}
\end{figure*}

In this section, we analyze the results of Model T-3D-0, which is the same as Model T-3D-2000 but restarts from the 2D simulation at $t=0$~years rather than $t=2000$~years. Since the disk has no initial substructure, it is initially stable to RWI. We expect the non-axisymmetric RWI modes to grow as rings and gaps develop through non-ideal MHD processes in the wind-launching disk. The initial development of nearly axisymmetric rings and gaps is illustrated in the top panels of Fig.~\ref{fig:RWI_Restart_2}, where we plot the normalized column density, the midplane density, plasma-$\beta_z$ based on the vertical component of the magnetic field, and radial velocity normalized by the local sound speed, at the same representative time of $t=340$~years as shown in the top panels for Model T-3D-2000 in Fig.~\ref{fig:RWI_Restart_1}. In the T-3D-2000 case, a prominent $m=8$ azimuthal RWI mode has already become clearly visible at this time in a ring around $\sim 11$~au radius. In the T-3D-0 case, although rings and gaps have developed near $\sim 10$~au at this time, they remain nearly axisymmetric, likely because their amplitudes are still relatively small and it takes time for RWI modes to grow, particularly on a background ring/gap of relatively low contrast. 

Non-axisymmetric RWI modes do develop at later times. For example, the middle row of Fig.~\ref{fig:RWI_Restart_2} shows a prominent $m=5$ mode on a ring of $\sim 9.7$~au radius at the second representative time of $t=680$~years, corresponding to about 22.5 local orbital periods. It is seen most clearly in the radial velocity map (Fig.~\ref{fig:RWI_Restart_2}h), but is also visible in the other three maps (Fig.~\ref{fig:RWI_Restart_2}efg). At later times, the non-axisymmetric structure becomes more dominated by lower-order azimuthal modes (with smaller $m$), broadly consistent with the trend expected for RWI. For example, at the end of the simulation ($t=1220$~years, shown in the bottom row), the azimuthal density perturbation is dominated by a high-density vortex ($m=1$ mode) near the bottom of the $\sim 9.7$~au-ring, although other low-order ($m=2$ and $m=3$) modes are also visible.

\begin{figure}
    \centering
    \includegraphics[width=\linewidth]{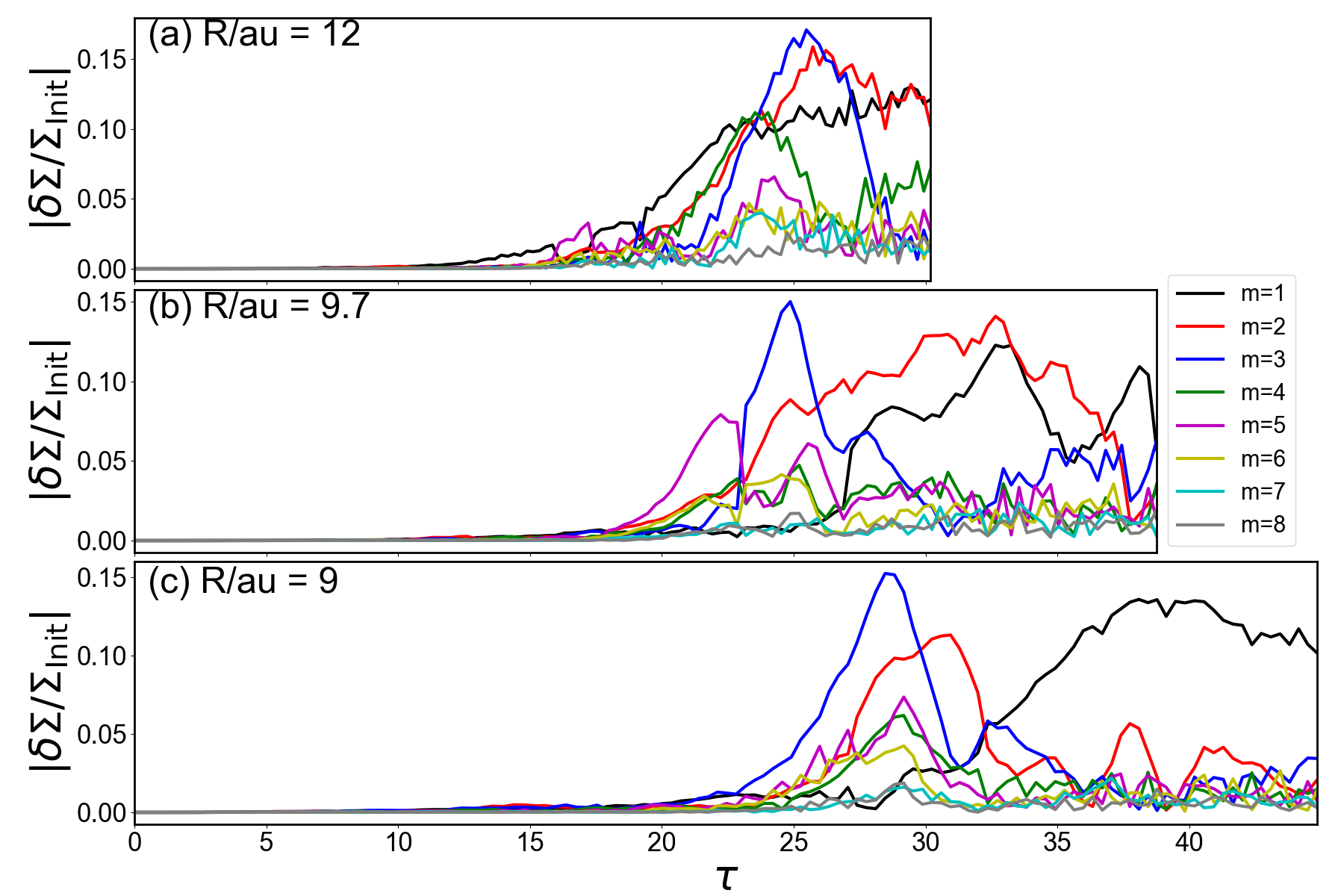}
    \caption{Same as Fig. \ref{fig:LateTimeModes}, but for Model T-3D-0 at R = 12 (a), 9.7 (b), and 9 au (c).  
    }
    \label{fig:LateTimeModes_T3D-0}
\end{figure}

To quantify the mode evolution, we plot in Fig.~\ref{fig:LateTimeModes_T3D-0} the time evolution of the amplitudes of various azimuthal modes, as done in Fig.~\ref{fig:LateTimeModes} for the restarted model T-3D-2000. It is clear that the $m=5$ mode initially dominates the $9.7$~au-ring (the purple peak around $\tau\sim 22$). As time progresses, it is first replaced by the $m=3$ (blue), then $m=2$ (red), and, finally, $m=1$ (black) mode. The progression towards lower-order modes with time is also evident for the $9$~au ring shown (bottom panel), consistent with the expectation of RWI. Interestingly, the top panel shows that the azimuthal modes of the outer $12$~au ring are dominated by $m=1$ at relatively early times between $\tau\sim 13$ - $16$. This is in contrast with the outer ($18$~au) ring of Model T-3D-2000, which is dominated by the high-order $m=8$ mode at similarly early times (see the gray curve in the top panel of Fig.~\ref{fig:LateTimeModes}). We attribute the different behavior to the difference in the initial disk structure when the 3D simulation starts. In the T-3D-2000 case, highly RWI unstable, well-defined rings of large contrast with their surroundings already exist at the beginning of the simulation (see the red curve in the top panel of Fig.~\ref{fig:RWI_2D}). For such (unstable) conditions, the higher-order modes are expected to grow faster, as discussed earlier in Section~\ref{subsec:3D_From_ReStart} (see also, e.g., \citealt{Ono18}). In the T-3D-0 case, there is no ring at 12~au (or anywhere else) initially. It takes time for the rings to form out of the initially smooth disk and for their amplitudes to increase relative to their surroundings (see Fig.~\ref{fig:SpaceTime_T3D-0} below). When the ring amplitude is smaller, the fastest-growing mode tends to be of a lower order \citep[see, e.g., Table 1 of][]{Ono18}, 
consistent with the above difference between the outer rings of Models T-3D-0 and T3D-2000. 

\begin{figure*}
    \centering
    \includegraphics[width=\linewidth]{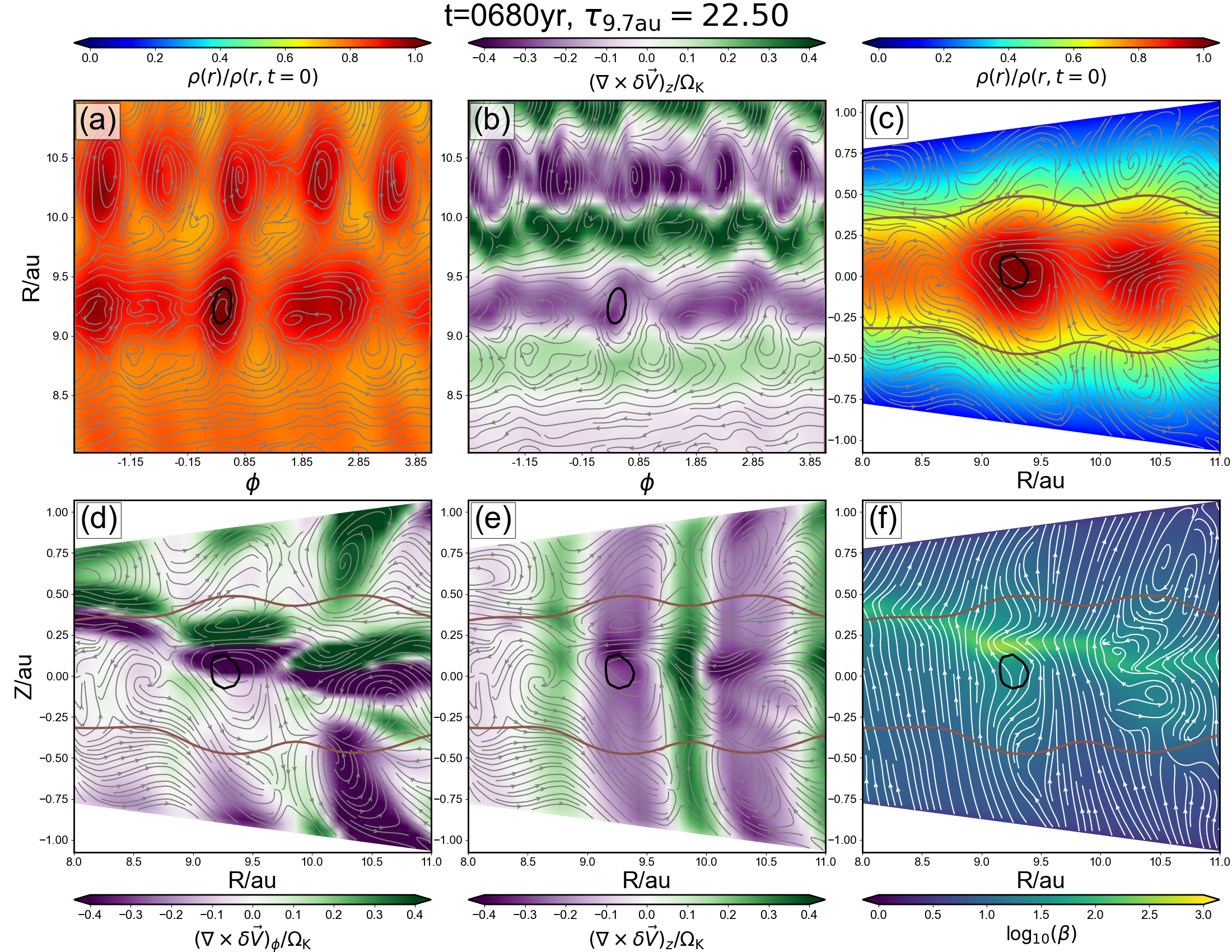}
    \caption{The structure of the dominant vortex on the 9.7~au ring in Model T-3D-0 at $\tau \sim 22.5$. Panels  (a) and (b) display the mass density normalized to its initial value at the midplane and the $z$ component vorticity normalized to its local Keplerian period, respectively. Panels (c)-(f) plot the mass density normalized to its initial value (c), the $\phi$ component of the vorticity normalized to its local Keplerian period (d), the $z$ component vorticity normalized to its local Keplerian period (e), and the logarithm of the plasma-$\beta$ (f) on a meridional plane passing through the density peak of the $\sim$ 9.2 au ring. The gray contours with arrows are velocity streamlines and the white contours (in panel [f]) with arrows are the magnetic field lines. The brown and black contour lines mark where the mass density normalized to its initial value is 0.6 and 1.0, respectively. 
    An animated version of the figure can be found at \url{https://virginia.box.com/s/oofe09pt2qv01wkk42vv8e0tzqg3oy7k}
    }
    \label{fig:Vortex_T3D-0}
\end{figure*}

To illustrate the vortices formed in the T-3D-0 model more pictorially, we plot in Fig.~\ref{fig:Vortex_T3D-0} the normalized density, vertical component of the vorticity, and the plasma$-\beta$ on the midplane and in the meridional plane passing through the center of a vortex at the same representative time of $t=680$~yrs as in the second row of Fig.~\ref{fig:RWI_Restart_2}, similar to the maps shown Fig.~\ref{fig:VortexStructure_1} for model T-3D-2000. Compared to the T-3D-2000 case, the density contrast between regions of different radii is much less (compare Fig.~\ref{fig:Vortex_T3D-0}a and Fig.~\ref{fig:VortexStructure_1}), which is expected given that the T-3D-2000 model was restarted with already formed high-contrast rings and gaps. Nevertheless, regularly spaced anti-cyclonic vortices (with negative values of vertical vorticity) are clearly visible at higher-density radii (e.g., $\sim 9.3$ and $10.3$~au) and cyclonic vortices (with positive values of vertical vorticity) at lower-density radii (e.g., $\sim 8.8$ and $9.8$~au; see panels [a] and [b]), with the cyclonic vortices forming preferentially at azimuthal angles between the anti-cyclonic ones (and vice versa). 

Despite the prominent vortices of both positive and negative vorticities on the $R-\phi$ plane, the flow structure in the meridional plane at this time is rather regular, with a layer of accreting material located approximately near the midplane, which drives two prominent cells of opposite meridional circulation above and below the layer near the radius of the dominant anti-cyclonic vortex marked by the black contour in panel (c). Indeed, associated with the meridional circulations is an azimuthal component of vorticity that is larger than the vertical component, even inside the dominant RWI vortex, as seen by comparing panels (d) and (e) of Fig.~\ref{fig:Vortex_T3D-0}. The strong meridional circulation, ultimately induced by the magnetic field through accretion and disk wind, is a feature absent from hydro simulations of RWI.  

The meridional flow pattern in 3D is nearly identical to that studied in detail in 2D (axisymmetric) simulations (e.g., Fig.8 of \citealt{Hu22}). In particular, the accretion layer is associated with regions of vanishing toroidal magnetic fields, where the field lines change directions sharply in the azimuthal direction, creating a large tension force in the (negative) azimuthal direction that brakes the rotation and drives accretion. The weak toroidal field is why the accretion layer shows up as a filament of high plasma$-\beta$ in panel (f). The panel shows clearly that the accretion drags the poloidal field lines into a pinched configuration, which, \cite{Suriano18} argue, leads to reconnection that lies at the heart of ring and gap formation in the first place. Clearly, the development of RWI in 3D does not shut off the mechanism for generating rings and gaps in the non-ideal MHD and wind-launching disk in the first place.  

\begin{figure*}
    \centering
    \includegraphics[width=0.9\linewidth]{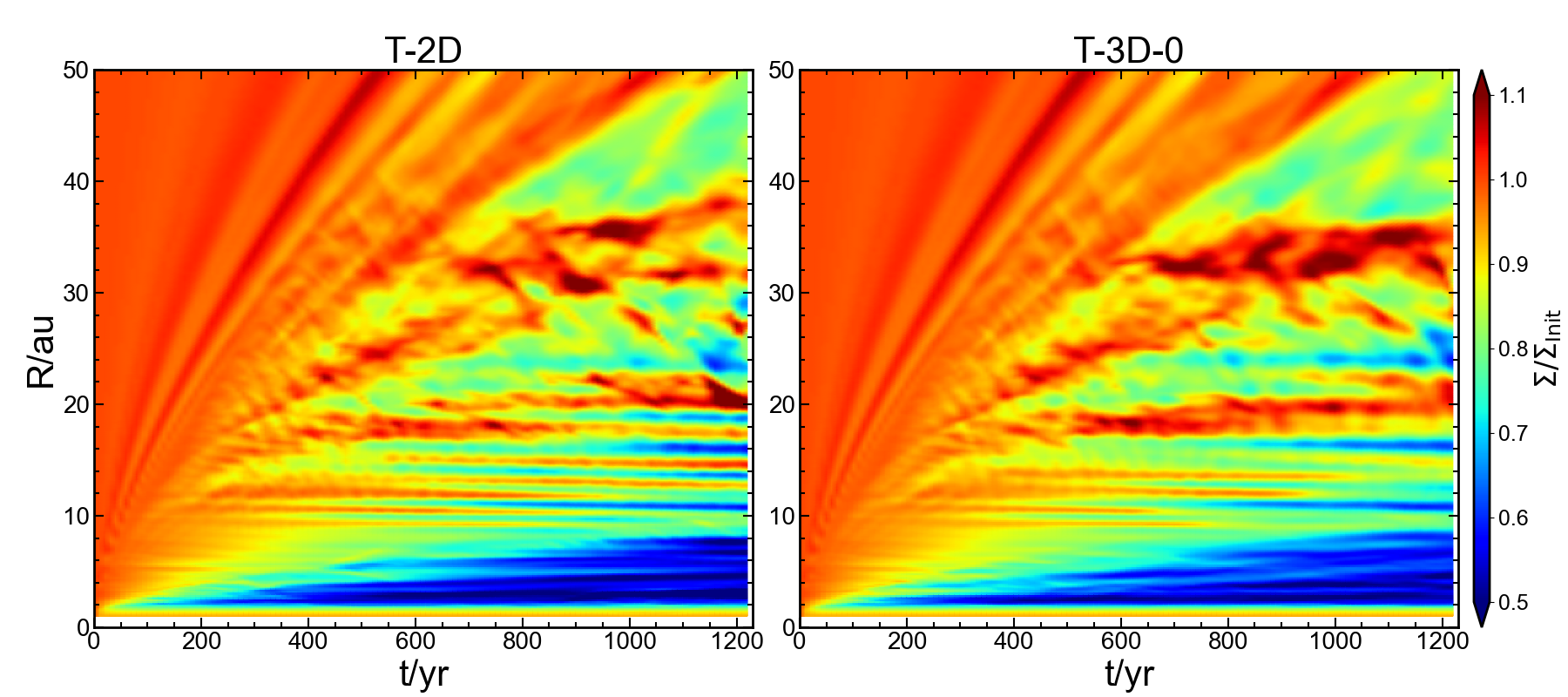}
    \caption{Same as the Fig. \ref{fig:RingGapSurvival}, but for the Model T-3D-0 and with a different color bar.
    }
    \label{fig:SpaceTime_T3D-0}
\end{figure*}

The persistence of ring and gap formation despite the RWI is further illustrated in Fig.~\ref{fig:SpaceTime_T3D-0}, where we compare the radial distribution of the azimuthally averaged normalized surface density at different times for Model T-3D-0 and its 2D counterpart T2D. As in the T-3D-2000 case (see Fig.~\ref{fig:RingGapSurvival}), the contrast between the rings and gaps tends to be somewhat less in 3D than in 2D. Nevertheless, rings and gaps have clearly developed and persisted until the end of the simulation. It can also be seen from the animated version of Fig.~\ref{fig:RWI_Restart_2} (see its caption for a link to the animation).

\begin{figure*}
    \centering
    \includegraphics[width=\linewidth]{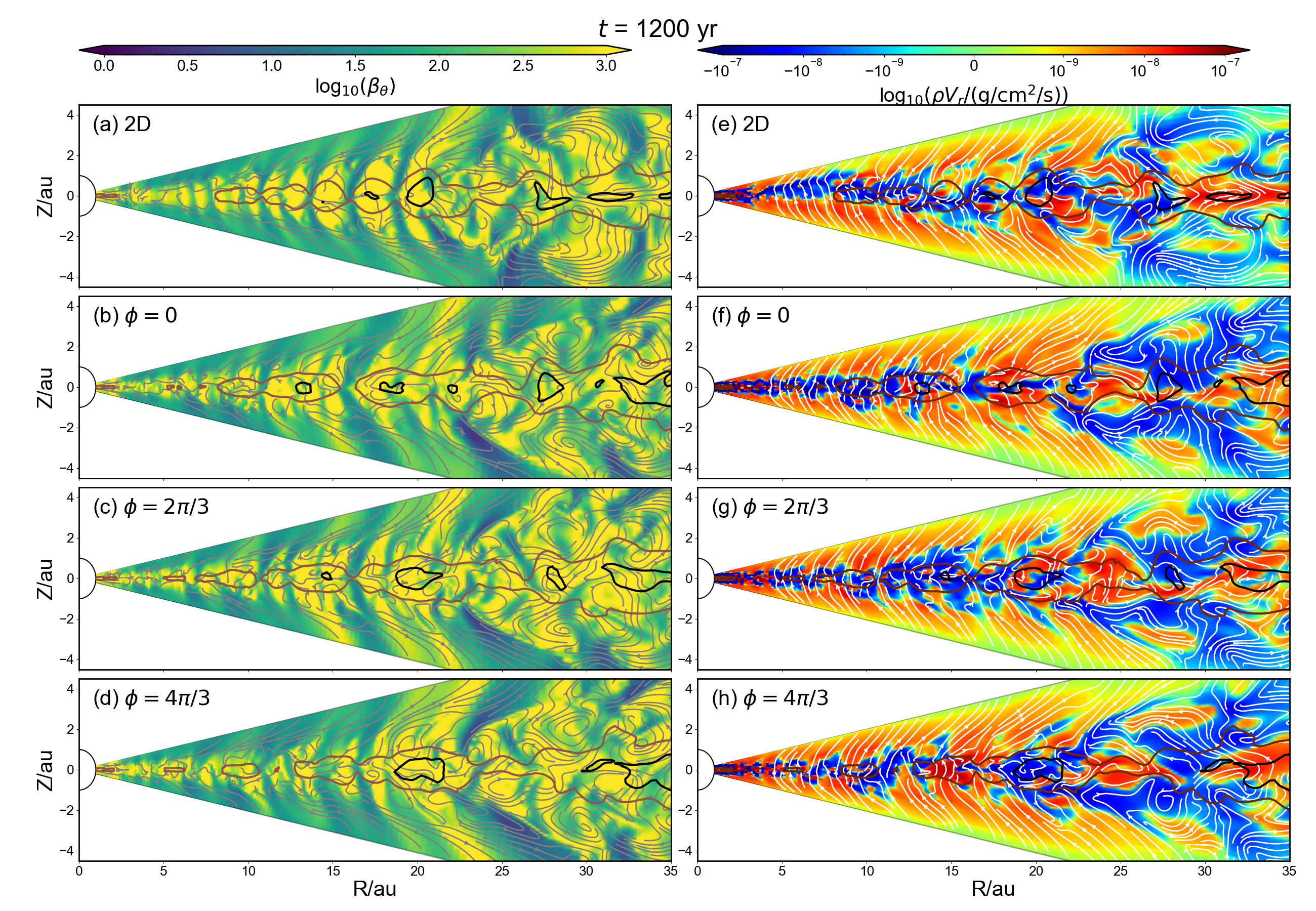}
    \caption{Comparison of the 2D and 3D meridional structures at a representative time $t=1200$~years. The left panels plot the logarithmic scale of plasma $\beta_\theta$ (based on the polar component of the magnetic field), with velocity streamlines traced by the gray lines with arrows. The right panels plot the radial mass flux per unit area, with the magnetic field lines traced by the white lines with arrows. The top panels are for the axisymmetric model T-2D, and the lower three panels are for three meridional planes of the 3D model T-3D-0 with, respectively, $\phi = 0$, $2\pi/3$, and$4\pi/3$. The brown and black contour lines mark where the mass density normalized to its initial value is 0.6 and 1.0, respectively. 
    }
    \label{fig:2D3D_Comp}
\end{figure*}

We believe the ring and gap formation persists because the RWI modifies but does not fundamentally change the flow structures that enable the formation and growth of rings and gaps. This is illustrated in Fig.~\ref{fig:2D3D_Comp}. The top left panel shows a much lower plasma-$\beta_\theta$ (based on the $\theta$-component of the magnetic field) in the low-density gaps than in the dense rings, consistent with the result from earlier work that the poloidal magnetic flux tends to be redistributed from the rings to the gaps. This poloidal magnetic flux redistribution pattern is fundamental to the MHD mechanism of ring and gap formation \cite[e.g.,][]{Suriano18,Riols19,Cui21}. It is broadly preserved in 3D, as seen in the three lower left panels, which plot the meridional distributions of the plasma-$\beta_\theta$ at three representative azimuthal angles. Also evident from the panels are azimuthal variations of the density (particularly in the rings between $\sim 10$ and $15$~au) and the plasma-$\beta_\theta$ (particularly around $\sim 16$~au where its value is much higher at $\phi=4\pi/3$ than at $\phi=0$ and $2\pi/3$). Similarly, the characteristic disk flow structure in 2D, with the accretion concentrating in warped layers (especially within about 20~au radius where most rings and gaps are located; see the blue regions in the top right panel), is broadly preserved in 3D. \cite{Suriano18} argued that this flow structure is important for ring and gap formation because it generates sharp pinching of the poloidal magnetic field, facilitating reconnection and magnetic flux redistribution relative to mass. There are azimuthal variations in the accretion flow, as seen in the three lower-right panels around, e.g., the $\sim$15-au radius. They appear to modify but not destroy the rings and gaps. 

A caveat of our simulations is that, despite the use of static mesh refinement, the resolution may not be high enough to fully resolve any MRI turbulence that may develop, especially in the relatively well (magnetically) coupled outer part of the disk beyond $\sim 40$~au, where the initial ambipolar Elsasser number on the midplane exceeds 10 for the 3D models discussed in this section (see the red curve in the upper panel of Fig.~\ref{fig_non_ideal_comparison}). Although our focus is on the less well-coupled inner disk (inside $\sim 30$~au) that is expected to be less prone to MRI than the outer disk, it is possible that its dynamics and structure can be modified at a higher resolution, where a better resolved MRI turbulence can potentially reduce the amplitudes of the rings and gaps and increase their widths, making them more stable to RWI. Indeed, there are no obvious vortices in the 3D model shown in Fig.~10 of Cui and Bai (2021), which has a higher resolution than ours but a smaller azimuthal angle range ($\pi/4$ rather than $2\pi$). Whether the limited azimuthal angle range has a strong effect on the formation and evolution of RWI vortices or not remains to be determined. Higher-resolution simulations with a full azimuthal angle range are desirable to resolve the potential discrepancy.

%% file: table_cases.tex
\begin{table}
\caption {
Models
\label{table_cases}
} 
\begin{tabular}{lll}
  Name  & Description  & Starting time $t_{\rm st}$\\
        \hline
    S-2D   &  Shu-type power-law ion abundance & 0 \\
    T-2D   &  chemical network   & 0 \\
    T-3D-2000  & chemical network    & 2000 years \\
    T-3D-0     & chemical network   & 0
\end{tabular}
\end{table}

%% file: Conclusion.tex

We performed 2D (axisymmetric) and fully 3D non-ideal MHD simulations of substructure formation in magnetized, weakly ionized, wind-launching disks, with the non-ideal MHD coefficients computed from a simplified chemical network including dust grains. A 2D simulation using a simple power-law prescription for the ionization fraction is also included for comparison. We focused on whether the substructures formed in such disks are stable to the Rossby Wave Instability (RWI) and, if not, how the RWI affects the formation and evolution of the substructures. Our main conclusions are summarized as follows. 

\begin{enumerate}
\item
In agreement with previous work, axisymmetric gas rings and gaps form spontaneously in 2D simulations of non-ideal MHD disks with ambipolar diffusion and Ohmic dissipation, with more prominent substructures for larger Elsasser numbers (or better magnetic coupling). The case for non-ideal MHD formation of substructures is thus strengthened, at least under the assumption of axisymmetry. 

\item Axisymmetric gas rings in 2D simulations are found to be unstable to Rossby Wave Instability (RWI) according to analytic stability conditions and through 3D restart simulation. As expected for RWI, shorter wavelength (or larger $m$) azimuthal modes develop earlier in the simulation, and longer wavelength ones dominate later, forming elongated anti-cyclonic vortices (arcs) in the ring's column density distribution that last until the end of the simulation. Highly elongated vortices with aspect ratios of 10 or more are found to decay with time in our simulation, in contrast with the hydro case. This difference could be caused by magnetically induced motions, particularly strong meridional circulations with large values of azimuthal vorticity, which may be incompatible with the columnar structure preferred by vortices. Axisymmetric gaps in the 2D simulation are also unstable to RWI, forming cyclonic vortices despite being strongly magnetized, with a plasma-$\beta$ of order unity. Nevertheless, the cyclonic and anti-cyclonic RWI vortices modify but do not destroy the rings and gaps in the radial gas distribution of the disk. 

\item 
Our 3D simulation that starts from a smooth initial condition shows that RWI does not shut off the mechanism for generating rings and gaps in the non-ideal MHD and wind-launching disk. Specifically, anti-cyclonic and cyclonic vortices are still formed in rings and gaps, respectively, with amplitudes saturating at moderate levels. The RWI and associated vortices modify but do not suppress the poloidal magnetic flux accumulation in low-density regions and the characteristic meridional flow patterns that are crucial to the ring and gap formation. 
\end{enumerate}

An interesting question to address in future investigations is how RWI affects the {\it dust} distribution and the disk appearance in continuum emission. Most observed disk dust substructures appear rather axisymmetric, although a small fraction shows non-axisymmetric features, including arcs, which are potentially vortices. The RWI vortices discussed in this paper open up the possibility of producing such dusty vortices in non-ideal MHD wind-launching disks. However, detailed dust treatment is needed to explore this possibility quantitatively. The current simulation will also benefit from adaptive (rather than static) mesh refinement, which will better resolve the vortices formed in the global 3D simulations. Higher resolution is also desirable to better resolve any MRI turbulence that may develop, which can modify the disk dynamics and structure, which, in turn, affects its stability to RWI.

%% file: appendix_Chemical_models.tex
\begin{figure}
    \centering
    \includegraphics[width=\linewidth]{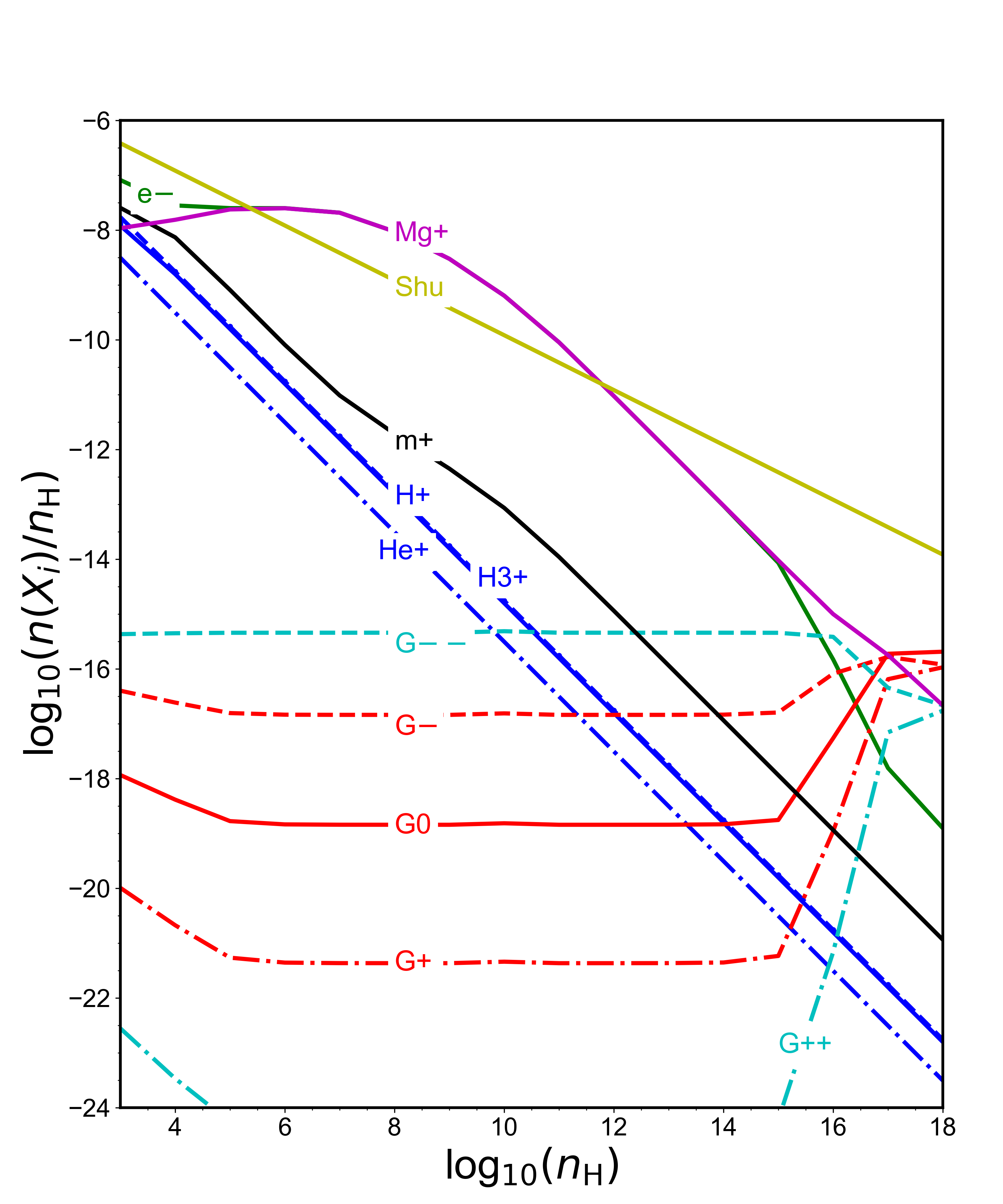}
    \caption{
    Fractional abundances of main charges as a function of the number density computed from the chemical network for an MRN-type dust size distribution with $a_{\rm min} = 0.5 \mu$m and $a_{\rm max} = 25 \mu$m. Magnesium ions, 
    molecular ions other than ${\rm H_3^+}$, grains with charge $0, \pm e$, and $\pm 2e$. The power-law prescription of \protect\cite{Shu92} is plotted as a yellow line for comparison. 
    }
    \label{fig_ion_with_Shu}
\end{figure}

One of our improvements over previous work along a similar line \citep{Suriano18, Hu22} is that we computed the charge abundances from a reduced chemical network (see \S~\ref{sec:non-ideal_coff}) rather than using the simple power-law prescription of \cite{Shu92}, which is derived under the assumption of one dominant molecular ion (e.g., ${\rm HCO^+}$) and no dust grains. 
Fig. \ref{fig_ion_with_Shu} shows fractional abundances of the main charges computed from our chemical network as a function of number density for an MRN-like dust size distribution with $a_{\rm min} = 0.5 \mu$m and $a_{\rm max} = 25 \mu$m,  including ${\rm Mg^+}$, ${\rm m^+}$ (molecular ions other than ${\rm H_3^+}$), grains with charge $0, \pm e$, and $\pm 2e$. The power-law prescription of \cite{Shu92} is also shown for comparison.

\begin{figure}
    \centering
    \includegraphics[width=\linewidth]{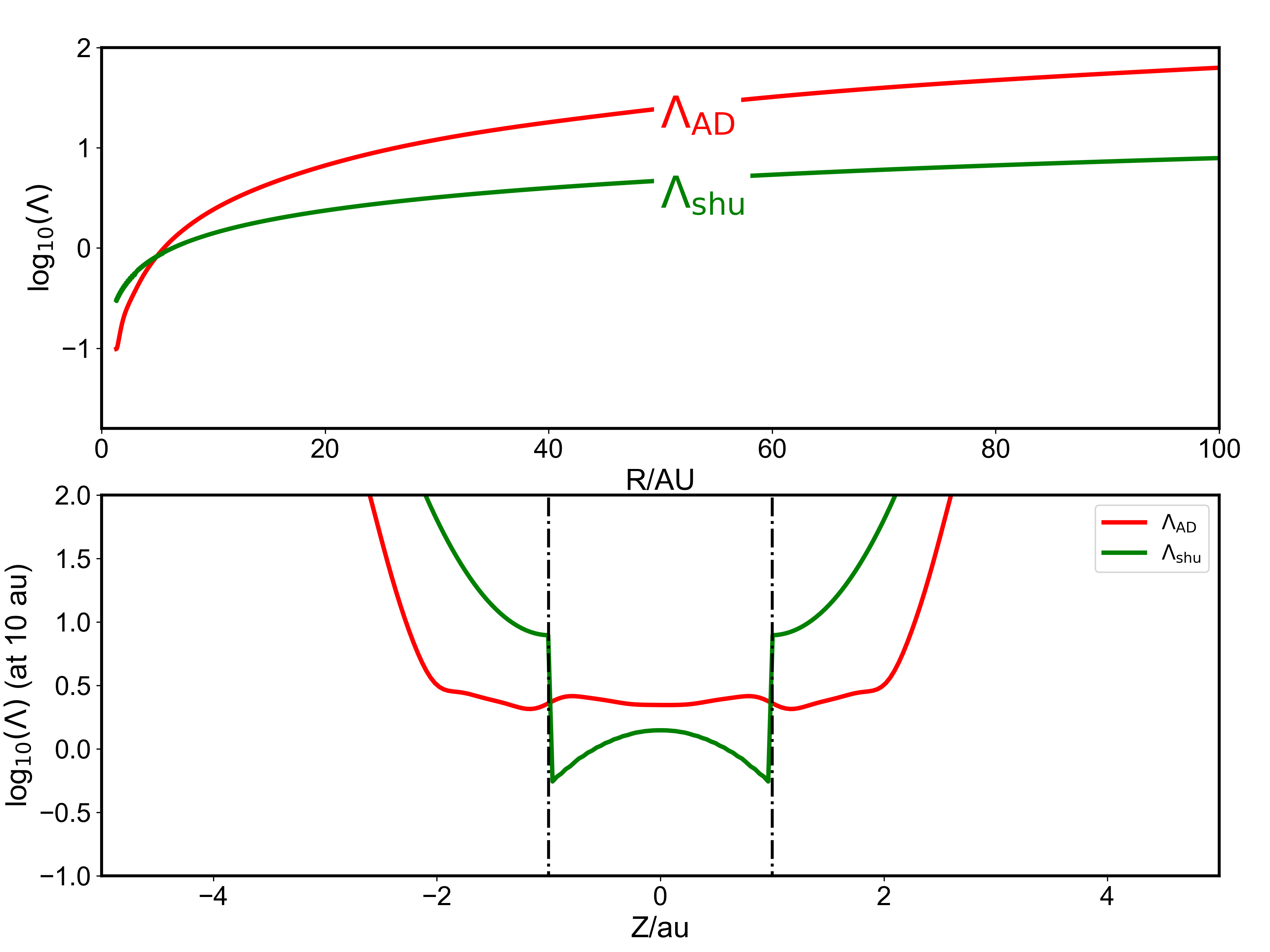}
    \caption{
    Initial profiles of the ambipolar Elsasser number for Model S-2D (green lines) and Model T-2D (red lines). Panel (a) shows the Elsasser numbers as a function of radius on the mid-plane. Panel (b) shows the Elsasser number at the radius $r=10$~au as a function of the vertical distance from the mid-plane. The dashed lines in panel (b) mark the disc surface near $\sim$ 1~au. 
    }
    \label{fig_non_ideal_comparison}
\end{figure}

We use the charge abundances in Fig.~\ref{fig_ion_with_Shu} to calculate the Elsasser number together with the disk properties. Fig. \ref{fig_non_ideal_comparison} shows the initial profile of the ambipolar Elsasser numbers of Model S-2D (where the power-law prescription of \cite{Shu92} is adopted; green lines) and Model T-2D (where the chemical network is used; red lines) as a function of radius on the mid-plane (panel a) and as a function of vertical distance from the midplane at $r=10$~au (panel b, where the $\theta$-dependence in equation~[\ref{eq:theta_depend_non_ideal}] is applied). 

\input{table_abundance}
\input{table_molecular_reaction}

%% file: table_abundance.tex
\begin{table}
\caption{
Relative abundances of the elements included in the chemical network. Similar to Table 1 of \protect\cite{Umebayashi90}, with the fractions remaining in the gas phase set to $\delta_1=0.2$ and $\delta_2=0.02$.}
\label{ch_table_abundance}
\begin{tabular}{llllll}
  Element  & Abundant & Chemical species & Fraction (gas) \\
        \hline
  ${\rm H}$    & 1                    & ${\rm H_2}$  & 1\\
  ${\rm He}$   & 8.5 $\times 10^{-2}$ & ${\rm He}$   & 1\\
  ${\rm C}$    & 4.2 $\times 10^{-4}$ & ${\rm CO}$   & $\delta_1$\\
  ${\rm O}$    & 6.9 $\times 10^{-4}$ &              &           \\
               & 9.0 $\times 10^{-5}$ & ${\rm O}$    & $\delta_1$\\
               & 9.0 $\times 10^{-5}$ & ${\rm O_2}$  & $\delta_1$\\
         \hline
  ${\rm Mg}$   & 1.7 $\times 10^{-4}$ & ${\rm Mg}$   & $\delta_2$\\
\end{tabular}
\end{table}

%% file: table_molecular_reaction.tex
\begin{table}
\caption {
List of gas-phase reactions in our chemical network. Symbols $\alpha_{\rm c}$, $\beta_{\rm c}$, and $\gamma_{\rm c}$ are the coefficients used to determine the reaction rates. These coefficients are grabbed from UMIST database. 
\label{ch_table_molecular_reaction}
} 
\begin{tabular}{llllll}
  Reaction  & $\alpha_{\rm c}$ & $\beta_{\rm c}$& $\gamma_{\rm c}$ \\
        \hline
  ${\rm H}^+ + {\rm O} \longrightarrow {\rm O}^+ + {\rm H}$ & $6.86 \times 10^{-10}$& 0.26 & 224.30\\
  ${\rm H}^+ + {\rm O_2} \longrightarrow {\rm O_2}^+ + {\rm H}$ & $2.00 \times 10^{-9}$& 0.0 & 0.0\\
  ${\rm H}^+ + {\rm Mg}   \longrightarrow {\rm Mg}^+   + {\rm H}$ &  $1.1 \times 10^{-9}$& 0.0 & 0.0\\
  
  ${\rm He}^+ + {\rm H_2}   \longrightarrow {\rm H}^+   + {\rm H} + {\rm He}$  &  $7.2 \times 10^{-15}$& 0.0 & 0.0\\
  ${\rm He}^+ + {\rm CO}    \longrightarrow {\rm C}^+   + {\rm O} + {\rm He}$  &  $1.6 \times 10^{-9}$& 0.0 & 0.0\\
  ${\rm He}^+ + {\rm O_2}   \longrightarrow {\rm O}^+   + {\rm O} + {\rm He}$  & $1.1 \times 10^{-9}$& 0.0 & 0.0\\

  ${\rm H_3}^+ + {\rm CO}   \longrightarrow {\rm HCO}^+   + {\rm H_2} $  & $1.36 \times 10^{-9}$& -0.14 & -3.40\\
  ${\rm H_3}^+ + {\rm O}    \longrightarrow {\rm OH}^+    + {\rm H_2} $  & $7.98 \times 10^{-10}$& -0.16 & 1.4\\
  ${\rm H_3}^+ + {\rm O_2}  \longrightarrow {\rm O_2H}^+  + {\rm H_2} $  & $9.3 \times 10^{-10}$& 0.0 & 100.0\\
  ${\rm H_3}^+ + {\rm Mg}    \longrightarrow {\rm Mg}^+     + {\rm H_2} + {\rm H}$  & $1.0 \times 10^{-9}$& 0.0 & 0.0\\

  ${\rm C}^+ + {\rm H_2}   \longrightarrow {\rm CH_2}^+   + h\nu $     & $2.0 \times 10^{-16}$& -1.3 & 23.0\\
  ${\rm C}^+ + {\rm O_2}   \longrightarrow {\rm CO}^+     + {\rm O} $  & $3.42 \times 10^{-10}$& 0.0 & 0.0\\
  ${\rm C}^+ + {\rm O_2}   \longrightarrow {\rm O}^+     + {\rm CO} $  & $4.54 \times 10^{-10}$& 0.0 & 0.0\\
  ${\rm C}^+ + {\rm Mg}     \longrightarrow {\rm Mg}^+     + {\rm C} $   & $1.1 \times 10^{-9}$& 0.0 & 0.0\\

  ${\rm O_2}^+ + {\rm Mg}     \longrightarrow {\rm Mg}^+     + {\rm O_2} $   & $1.2 \times 10^{-9}$& 0.0 & 0.0\\
  
  ${\rm HCO}^+ + {\rm Mg}     \longrightarrow {\rm Mg}^+     + {\rm HCO} $   & $2.9 \times 10^{-9}$& 0.0 & 0.0\\
  ${\rm H}^+ + {\rm e} \longrightarrow {\rm H} + h\nu$ & $3.5 \times 10^{-12}$& -0.75 & 0.0\\
  ${\rm He}^+ + {\rm e} \longrightarrow {\rm He} + h\nu$ & $5.36 \times 10^{-12}$& -0.50 & 0.0\\
  
  ${\rm H_3}^+ + {\rm e} \longrightarrow {\rm H} + {\rm H} + {\rm H}$ & $4.36 \times 10^{-8}$& -0.52 & 0.0\\
  ${\rm H_3}^+ + {\rm e} \longrightarrow {\rm H_2} + {\rm H} $ & $2.34 \times 10^{-8}$& -0.52 & 0.0\\

  ${\rm C}^+ + {\rm e} \longrightarrow {\rm C} +  h\nu $ & $2.36 \times 10^{-12}$& -0.29 & -17.6\\
  ${\rm Mg}^+ + {\rm e} \longrightarrow {\rm Mg} +  h\nu $ & $2.78 \times 10^{-12}$& -0.68 & 0.0\\

  ${\rm CO}^+ + {\rm e} \longrightarrow {\rm O} +  {\rm C} $ & $2.0 \times 10^{-7} $& -0.48 & 0.0\\
  
  ${\rm O_2}^+ + {\rm e} \longrightarrow {\rm O} +  {\rm O} $ & $1.95 \times 10^{-7} $& -0.70 & 0.0\\
  ${\rm OH}^+ + {\rm e} \longrightarrow {\rm O} +  {\rm H} $ & $3.75 \times 10^{-8} $& -0.50 & 0.0\\
  ${\rm O_2H}^+ + {\rm e} \longrightarrow {\rm O_2} +  {\rm H} $ & $3.0 \times 10^{-7} $& -0.50 & 0.0\\
  ${\rm HCO}^+ + {\rm e} \longrightarrow {\rm CO} +  {\rm H} $ & $2.4 \times 10^{-7} $& -0.69 & 0.0\\
  ${\rm CH_2}^+ + {\rm e} \longrightarrow {\rm C} +  {\rm H_2} $ & $7.68 \times 10^{-8} $& -0.60 & 0.0\\

\end{tabular}
\end{table}